\documentclass[12pt,review]{elsarticle}

\usepackage{lineno,hyperref}
\usepackage{subfigure}
\modulolinenumbers[5]

\journal{}

\usepackage{tabularx}
\usepackage{multirow}
\usepackage{graphicx}
\usepackage{tcolorbox}
\usepackage{float}
\usepackage{booktabs}
\usepackage{setspace}

\usepackage{soul}

\raggedright% flush left alignment

\newcommand{\revis}[1]{{\color{black} #1}}

\usepackage[tmargin=1in,bmargin=1in,lmargin=1in,rmargin=1in]{geometry}
\begin{document}

\begin{frontmatter}

\title{Exploring User Perspectives on ChatGPT: Applications, Perceptions, and Implications for AI-Integrated Education}

\author[1,2]{Reza Hadi Mogavi}
\ead{rhadimogavi@sigchi.org}

\author[3]{Chao Deng}
\ead{cdeng@accessiblemeta.org}

\author[1]{Justin Juho Kim}
\ead{jjkimab@connect.ust.hk}

\author[4]{Pengyuan Zhou}
\ead{pyzhou@ustc.edu.cn}

\author[5]{Young D. Kwon}
\ead{ydk21@cam.ac.uk}

\author[6,2]{Ahmed Hosny Saleh Metwally}
\ead{ahmed.hosny@bnu.edu.cn}

\author[6]{Ahmed Tlili}
\ead{ahmed.tlili23@yahoo.com}

\author[7]{Simone Bassanelli}
\ead{simone.bassanelli@unitn.it}

\author[7]{Antonio Bucchiarone}
\ead{bucchiarone@fbk.eu}

\author[8]{Sujit Gujar}
\ead{sujit.gujar@iiit.ac.in}

\author[2]{Lennart E. Nacke}
\ead{lennart.nacke@acm.org}

\author[1,9,10]{Pan Hui\corref{mycorrespondingauthor}}
\ead{panhui@ust.hk}

\cortext[mycorrespondingauthor]{Corresponding author}

\address[1]{The Hong Kong University of Science and Technology, Hong Kong SAR}
\address[2]{University of Waterloo, Canada}
\address[3]{Accessible Meta Group, United States}
\address[4]{University of Science and Technology of China, China}
\address[5]{University of Cambridge, United Kingdom}
\address[6]{Smart Learning Institute of Beijing Normal University, China}
\address[7]{Fondazione Bruno Kessler, Italy}
\address[8]{International Institute of Information Technology (Hyderabad), India}
\address[9]{The Hong Kong University of Science and Technology (Guangzhou), China}
\address[10]{University of Helsinki, Finland}

\tnotetext[mytitlenote]{This is the authors' preprint version of the paper accepted by the Journal of Computers in Human Behavior: Artificial Humans (doi: https://doi.org/10.1016/j.chbah.2023.100027)}

% %% Group authors per affiliation:
% \author{Elsevier\fnref{myfootnote}}
% \address{Radarweg 29, Amsterdam}
% \fntext[myfootnote]{Since 1880.}

% %% or include affiliations in footnotes:
% \author[mymainaddress,mysecondaryaddress]{Elsevier Inc}
% \ead[url]{www.elsevier.com}

% \author[mysecondaryaddress]{Global Customer Service\corref{mycorrespondingauthor}}
% \cortext[mycorrespondingauthor]{Corresponding author}
% \ead{support@elsevier.com}

% \address[mymainaddress]{1600 John F Kennedy Boulevard, Philadelphia}
% \address[mysecondaryaddress]{360 Park Avenue South, New York}

\begin{abstract}
\revis{To foster the development of pedagogically potent and ethically sound AI-integrated learning landscapes, it is pivotal to critically explore the perceptions and experiences of the users immersed in these contexts. In this study, we perform a thorough qualitative content analysis across four key social media platforms. Our goal is to understand the user experience (UX) and views of early adopters of ChatGPT across different educational sectors. The results of our research show that ChatGPT is most commonly used in the domains of higher education, K-12 education, and practical skills training. In social media dialogues, the topics most frequently associated with ChatGPT are \textit{productivity}, \textit{efficiency}, and \textit{ethics}. Early adopters' attitudes towards ChatGPT are multifaceted. On one hand, some users view it as a transformative tool capable of amplifying student self-efficacy and learning motivation. On the other hand, there is a degree of apprehension among concerned users. They worry about a potential overdependence on the AI system, which they fear might encourage superficial learning habits and erode students' social and critical thinking skills. This dichotomy of opinions underscores the complexity of Human-AI Interaction in educational contexts. Our investigation adds depth to this ongoing discourse, providing crowd-sourced insights for educators and learners who are considering incorporating ChatGPT or similar generative AI tools into their pedagogical strategies.}
\end{abstract}

\begin{keyword}
Artificial Intelligence (AI), Generative AI \sep ChatGPT \sep Education \sep Human-Computer Interaction (HCI) \sep  \sep Early Adopters  \sep Social Media \sep Qualitative Research
% \texttt{elsarticle.cls}\sep \LaTeX\sep Elsevier \sep template
% \MSC[2010] 00-01\sep  99-00
\end{keyword}

\end{frontmatter}

% \linenumbers

\section{Introduction} \label{sec:introduction}
\revis{The twenty-first century is characterized by a rapid proliferation of Artificial Intelligence (AI) technologies, leading to profound transformations across various aspects of our daily lives, a theme explored in numerous studies (e.g., see \cite{10.1145/3359313, Aktan2022, Wang2023, 10.1145/3396956.3398260, Zhai2021, 10.1145/3539597.3575784}). Among these transformative changes, the most significant is arguably the revolutionary impact on the educational sector \cite{Zhai2021, seldon2018fourth}.

This impact is best exemplified by the introduction of AI chatbots in education, which provide instantaneous responses to students' inquiries, thus serving as readily accessible learning resources at any time and location \cite{Clarizia2018, Hwang2021, Kasthuri2021}. Concurrently, the use of cutting-edge machine learning algorithms to analyze student data has empowered educators to pinpoint areas where individual students require additional support, thereby enhancing the efficacy of teaching strategies \cite{Zhang2022}. Furthermore, the advent of AI-based learning systems has paved the way for adaptive learning. These systems, designed to adjust to each student's learning pace and style, provide bespoke educational content. Such personalization has demonstrated its potential in improving learning outcomes and engagement levels \cite{10.1145/3406865.3418326, HadiMogavi2021, 10.1145/3290605.3300864}.

The Chat Generative Pre-trained Transformer (ChatGPT)\footnote{\url{https://chat.openai.com/}} is an advanced chatbot technology that utilizes Natural Language Processing (NLP) to generate coherent and contextually-relevant responses to a variety of human inquiries. In accordance with formal definitions, a chatbot is a ``computer program designed to conduct a conversation with a user through text or speech'' \cite{Chetlen2019}. The ability of ChatGPT to create conversational and engaging learning environments, customized to individual student needs, underscores its potential for application in educational contexts. This potential is increasingly recognized, as evidenced by an exponential growth in global interest in recent months, a trend observable in worldwide Google search statistics (refer to Figure~\ref{fig:statista}).

This surge in interest highlights the promising prospects of ChatGPT for the development of advanced AI-powered educational systems. However, it simultaneously underscores the need for a discerning approach when introducing this technology into educational settings. The caution advised in its usage stems from potential risks that have been identified in existing research \cite{sok2023chatgpt}. As such, the broadening interest in ChatGPT should be tempered by a careful evaluation of these potential risks.

\begin{figure*}[]
    \centering
\includegraphics[width=0.6\textwidth,keepaspectratio]{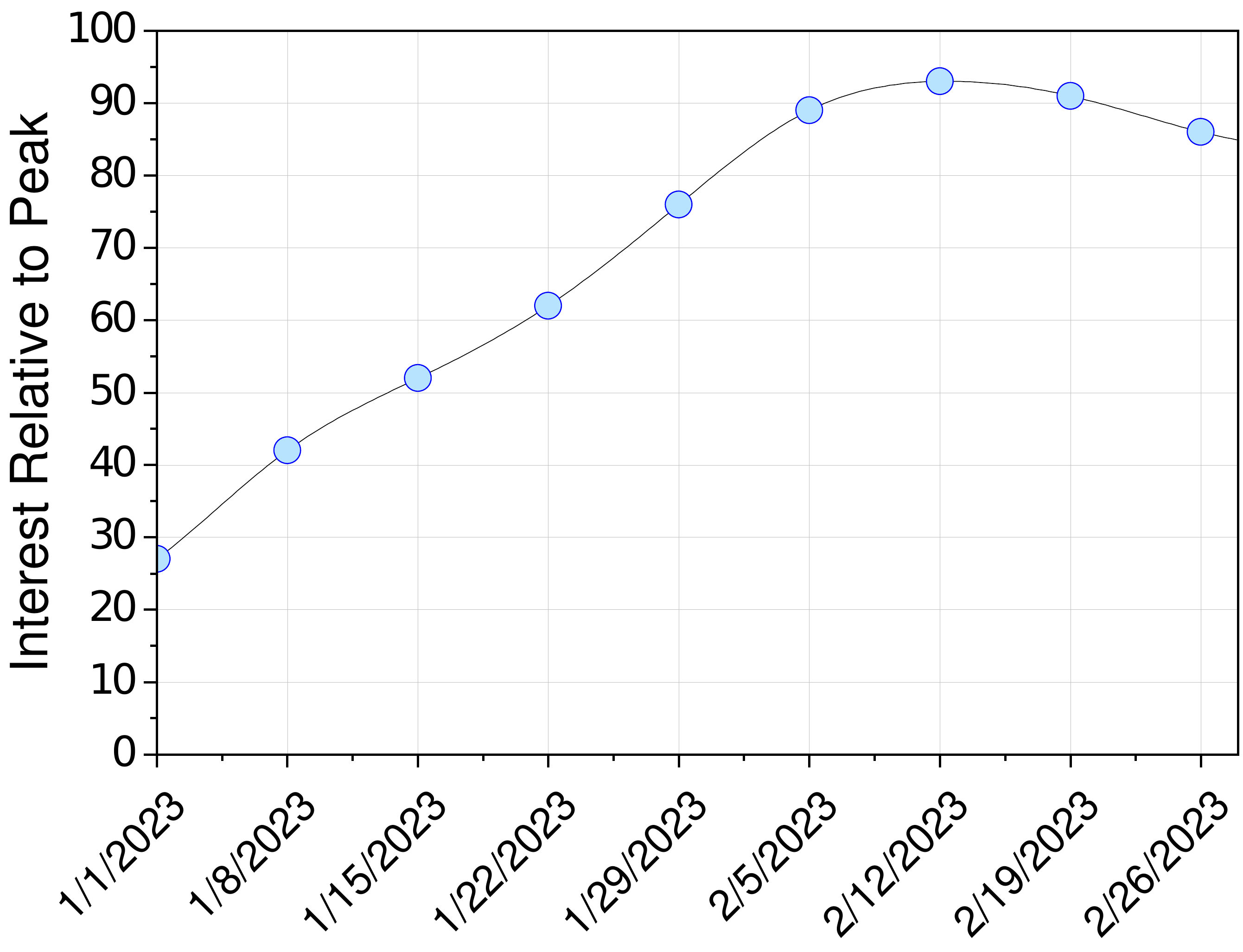}
    \vspace*{-5mm}
    \caption{The weekly search interest in ChatGPT on Google has consistently increased since its inception, according to Statista \cite{MyStatista}.}
    \label{fig:statista}
\end{figure*}

\textbf{Why Turn to Social Media for Insights?} Since its official rollout, ChatGPT has rapidly accumulated a significant user base \cite{gptusers2023}. Notably, social media has emerged as a primary venue for users to share their experiences and viewpoints regarding this AI tool \cite{Tlili2023}. These publicly articulated opinions on social media platforms can yield valuable insights into the usage of ChatGPT \cite{Tlili2023}. This data enables researchers to delineate the primary educational applications of ChatGPT and to assess its potential impacts, both positive and negative, at an incipient stage.% Given this context, it would be beneficial to conduct an initial empirical study to address these questions.

With this background in mind, we are motivated to take the first steps to fill this empirical research gap systematically by conducting a social media content analysis (SMCA) of four prominent social media platforms: \textit{Twitter}\footnote{https://twitter.com/}, \textit{Reddit}\footnote{https://www.reddit.com/}, \textit{YouTube}\footnote{https://www.youtube.com/}, and \textit{LinkedIn}\footnote{https://www.linkedin.com/}. 

SMCA is a popular research approach in the field of Human-Computer Interaction (HCI) that allows researchers to capture people's reflective responses \cite{10.1145/3491140.3528274, smith1992motivation} to an emerging technology on a demographically large scale, which can be otherwise quite difficult to establish through interviews, surveys, or other intrusive research methods \cite{10.1145/3491140.3528274, Sinha2021, Blandford2016}.
% \footnote{The selection of platforms for our study was based on their popularity, diverse user base, favorable terms of use, and potential to generate informative and varied data \cite{Sinha2021, Blandford2016}.}

\textbf{Research Questions.} More specifically, our research in this paper is guided by two main research questions:
\begin{itemize}
    \item \textbf{RQ1:} What are the predominant educational applications of ChatGPT as reported by early adopters?
    % Different people in different learning sectors use the technology differently
    \item \textbf{RQ2:} What are the initial perceptions of adopters within the education sector towards ChatGPT, and how do they evaluate its impact on the educational process? (Is ChatGPT a blessing or a curse?)
    % Anxiety, apprehension, etc.
\end{itemize}
}
In our research, we consider the perspectives of various early adopter groups, including \textit{researchers}, \textit{educators}, \textit{learners}, \textit{parents}, and \textit{general users (unspecified roles)}, all of whom could potentially play a critical role in improving our understanding of the development, implementation, and use of educational technology \cite{10.1145/3430895.3460126, Tlili2023, Tlili2022, Cheng2014}. \revis{We posit that such a comprehensive investigation can inform the inclusive and responsible advancement of ChatGPT in education, especially considering its nascent stage of adoption and development. In what follows, we outline some of the primary findings of our research:}

% \subsection{Key Findings}
\textbf{ChatGPT in Education.} Utilizing an inductive coding approach, this study scrutinized approximately 150,000 words contained in 6,000 text samples to ascertain the prevalent applications of ChatGPT across various educational contexts. A noteworthy adoption of ChatGPT was observed in higher education (24.18\%), K-12 education (22.09\%), and practical skills learning (15.28\%). Additionally, ChatGPT has found significant use in professional development (9.61\%), entrepreneurship (7.18\%), lifelong learning (4.64\%), and popular science education (3.25\%). A total of 13.77\% of the instances pertained to unclassified learning areas.

ChatGPT's versatility of use was demonstrated in its wide array of applications including, but not limited to, assisting in essay and paper composition, guiding learners through homework assignments and test preparation, aiding in language acquisition, fostering career progression, honing communication strategies, improving digital literacy, developing management competencies, facilitating benchmarking and brainstorming sessions, supporting decision-making processes, providing microlearning experiences, and simplifying complex concepts in popular science education. Further details are elaborated in Section \ref{sec:findings}.

\textbf{ChatGPT: A Double-Edged Sword.} \revis{Discussions about ChatGPT on social media platforms primarily focus on three themes: \textit{productivity} (73.19\%), \textit{efficiency} (63.02\%), and \textit{ethics} (48.51\%).\footnote{Readers should note that these percentages are not mutually exclusive; some user quotes may fall into multiple categories.} Our research suggests that ChatGPT can have both positive and negative implications for productivity, efficiency, and ethics.

Pioneer users of ChatGPT in educational contexts have expressed varied perceptions. Specifically, 30.62\% reported solely positive experiences, 17.15\% offered only negative feedback, 37.34\% expressed mixed feelings, and 14.89\% maintained a neutral stance.

On one hand, ChatGPT is seen as a helpful tool that enhances students' self-efficacy and learning motivation. This is particularly beneficial for students with special needs and/or everyone else who might struggle with traditional teaching methodologies. The conversational style of ChatGPT enriches the learning experience. Its capacity to provide immediate feedback and personalized learning opportunities fosters a deeper connection between students and the subject matter, leading to enhanced knowledge retention.

Furthermore, the round-the-clock availability of ChatGPT is a boon for students with demanding schedules, ensuring they do not miss out on learning opportunities. This is especially important for students from socioeconomically disadvantaged backgrounds who may have commitments that conflict with traditional learning schedules.

On the other hand, some concerns exist. There are apprehensions that students might rely excessively on AI systems, leading to superficial learning, reduced social skills, and impaired communication abilities. It is also feared that it might hinder the development of key psychological traits and discourage exploration of alternative viewpoints.

In addition, the deployment of ChatGPT in education raises vital concerns about data privacy and ethics. Further details are elaborated in Section \ref{sec:findings}.}

% \subsection{Main Contributions}
\textbf{Contributions.} In summation, this study offers several primary contributions to the field:
\begin{itemize}
    \item We have developed a comprehensive resource that catalogues the most common applications of ChatGPT within educational environments. 
    \item To the best of our knowledge, our study is the first large-scale qualitative investigation based on actual user feedback that explores the viewpoints of early adopters of ChatGPT in the educational context. By illuminating the diverse array of perspectives held by early adopters in the education sector, we provide a nuanced understanding of the positive and negative perceptions surrounding ChatGPT. 
    \item The findings from our investigation reveal that ChatGPT is not designed to supplant the roles of teachers; instead, it is supposed to function as an auxiliary tool that assists them in elevating the quality of education. The support from educators is crucial for contextualizing and authenticating the content produced by ChatGPT. 
    % \item This research uses crowd-sourced knowledge of early adopters to provide recommendations for educators and learners.
    % \item Finally, the thematic analysis presented in our study can provide valuable insights for the development of future surveys, interviews, and other research methods (e.g., controlled experiments) that seek to comprehend the varied and extensive perspectives and experiences of diverse stakeholders in adopting ChatGPT in education.
    \item Finally, the key findings from our study can guide the design of future studies. These could include surveys, interviews, and controlled experiments that aim to better understand the diverse and comprehensive views and experiences of different stakeholders when incorporating ChatGPT in education.
\end{itemize}
%%
% \section{Related Work} \label{sec:related_work}
\section{Background and Related Work} \label{sec:related_work}
\subsection{Language Models}
The field of natural language processing, also known as computational linguistics, has made significant progress in recent years, especially with the emergence of large neural network models trained on massive datasets \cite{Zhong2021, 10.1145/3560815}. Unlike traditional rule-based systems, these large data-driven models achieve greater precision and efficiency in various NLP tasks \cite{10.1145/3491101.3503719, Esmaeilzadeh2021}, such as language translation \cite{https://doi.org/10.48550/arxiv.2002.07526}, sentiment analysis \cite{Zhang2018}, named entity recognition \cite{10.1145/3366423.3380127}, and text summarization \cite{Elsaid2022}, among others.

Deep learning has become the dominant approach for developing NLP models \cite{Lauriola2022}. Among the most popular NLP architectures today are recurrent neural networks (RNNs) \cite{Baktha2017}, which include long short-term memory networks (LSTMs) and gated recurrence units (GRUs) \cite{Yao2018, Baktha2017}, as well as state-of-the-art transformer models \cite{Wolf2020}. In particular, the transformer model introduced by Vaswani et al. \cite{vaswani_attention_2017} has been highly influential in the field, serving as the foundation for many large language models such as the GPT family \cite{Wolf2020}.

Transformers use a \textit{self-attention} mechanism to process input tokens in parallel, as opposed to sequentially, which is a key difference from recurrent neural networks (RNNs) \cite{vaswani_attention_2017, Wolf2020}. This allows transformers to be more efficient in processing large amounts of text data and capturing long-range dependencies between words. The self-attention mechanism enables the model to better understand relationships between words regardless of their position in the text, which improves performance on NLP tasks \cite{Wolf2020}.

Language models have greatly benefitted from the use of \textit{pre-training} techniques \cite{NEURIPS2019_c20bb2d9}. These techniques involve initially training a model on an extensive dataset before fine-tuning it for a specific task. Pre-trained models such as BERT (Bidirectional Encoder Representations from Transformers) \cite{devlin2018bert}, BART (Bidirectional and Auto-Regressive Transformers) \cite{lewis2019bart}, XLNet \cite{NEURIPS2019_dc6a7e65}, and LLaMa from Facebook/Meta \cite{touvron2023llama} have showcased significant improvements in performance across various natural language processing tasks.

Most large language models today possess the capability to acquire novel tasks with minimal examples, even if they are not directly associated with their initial training \cite{beltagy2022zero}. This phenomenon is called ``few-shot learning,'' which denotes the ability of these models to adapt quickly and efficiently to new tasks. This proficiency is derived from the extensive general knowledge they acquire during the initial training phase, which can be applied to novel tasks \cite{beltagy2022zero}. This transfer learning phenomenon has been witnessed in many contexts, including the creation of sophisticated chatbots and language translation models \cite{Kulkarni2021, Omran2023}.

\subsection{GPT Family Models}
The GPT family of models, Generative Pre-trained Transformers, introduced by OpenAI, has experienced significant advancements since the initial release of GPT-1 in 2018 \cite{radford2018improving, radford2019language, NEURIPS2020_1457c0d6, bubeck2023sparks}. These powerful LLMs have demonstrated proficiency in a wide range of NLP tasks and have grown steadily in size and complexity \cite{bubeck2023sparks, bang2023multitask}.

The pioneering model in the GPT series, GPT-1 \cite{radford2018improving}, was built on the transformer architecture and boasted 117 million parameters, allowing it to process vast amounts of textual information. This inaugural model established a solid groundwork for subsequent advancements in the GPT series by laying the foundation for the exploration and enhancement of transformer-based language models. Despite representing a noteworthy milestone in natural language processing, GPT-1 exhibited certain shortcomings. For instance, the model tended to generate redundant text, particularly in response to prompts that were beyond the purview of its training data. Moreover, GPT-1 struggled to reason in multiple exchanges within a dialogue and was unable to maintain long-term dependencies within the textual content. Furthermore, while the model exhibited cohesion and fluency in short text sequences, these qualities diminished in longer passages.

In February 2019, OpenAI debuted GPT-2 \cite{radford2019language}, a language model that represented a notable advancement over its predecessor, boasting 1.5 billion parameters. This model served as a clear demonstration of the potential for LLMs to consistently generate text that is both cogent and contextually appropriate. However, it also raised concerns about the possibility of its malicious misuse (e.g., fake news generation) \cite{MITReviewgpt2}. Despite its numerous merits, GPT-2 was not without limitations, as it struggled with tasks that required more complex reasoning and contextual understanding.

GPT-3 \cite{NEURIPS2020_1457c0d6}, launched in June 2020, established a new standard in the field with an unparalleled 175 billion parameters. This model demonstrated an impressive ability to generate human-like text, enticing researchers and industries to explore its applications in various NLP tasks \cite{Dale2020, Abdullah2022}. The GPT-3.5 version further refined the capabilities of the model, which was later used for the development of ChatGPT.

The most recent addition to the GPT series, GPT-4 \cite{openai2023gpt4}, was unveiled on March 14, 2023, and has received widespread attention for its exceptional performance and multimodal capabilities. Estimated to have trillions of parameters, GPT-4 can efficiently process both textual and visual inputs, expanding its potential applications and pushing the limits of AI-generated content even further. Although currently accessible in a restricted capacity, GPT-4 marks a significant advancement in the field of large language models.

The evolution of the GPT family of models has had a profound impact on deep learning and NLP research \cite{zhang2023complete}. With each new iteration, these models have demonstrated increasingly sophisticated language generation capabilities, opening up new research avenues and applications in various domains \cite{radford2018improving, radford2019language, NEURIPS2020_1457c0d6, bubeck2023sparks, zhang2023complete}.

The GPT family is expected to continue to evolve with the release of GPT-5, anticipated in early 2024. This development is likely to bring about advances in language understanding, multimodal processing, and task-specific fine-tuning. These improvements are expected to contribute significantly to the ongoing progress in NLP and AI, driving transformative changes in research and industry contexts. However, as the capabilities of the GPT family grow, concerns about potential misuse and ethical implications \cite{Ray2023, zhang2023complete} are also expected to increase, especially in the context of education. Balancing the benefits and risks associated with these powerful LLMs will be paramount to ensure their responsible deployment and adoption.
\subsection{AI in Education}
\revis{AI's expansive influence across various sectors has had profound impacts on organizations, societies, and individuals \cite{Dwivedi2023}. Dwivedi et al. delved into these implications, examining the opportunities and challenges arising from generative AI technologies such as ChatGPT in education, business, and society \cite{Dwivedi2023}. They underscored AI's potential to enhance ``productivity'' across diverse sectors, while also drawing attention to issues such as privacy threats and misinformation.

Building on this, Calisto et al. investigated the acceptance and application of AI systems, with a particular focus on healthcare settings. They noted that health-focused AI tools like BreastScreening-AI were favorably received by target users, contributing to reduced diagnosis times and fewer errors \cite{Calisto2022, Calisto2023}. However, they emphasized that trust and acceptance from end-users are pivotal for the successful implementation of AI techniques \cite{Calisto20221}.

In the realm of education, AI's role has been predominantly about creating personalized, adaptive learning systems to enhance student experiences \cite{Holmes2022}. The shift in attention has gone from replacing to empowering teachers, emphasizing their crucial roles when deploying AI technologies in education \cite{Khosravi2022}. This is further supported by Shoufan's research on ChatGPT's usefulness for programming tasks, which highlights that AI cannot fully supplant human intelligence, as evidenced by the unavoidable errors in AI responses \cite{Shoufan2023}.

A cross-cultural study by Luo et al. provided insights from experts in China and the United States on the implications of ChatGPT for early education \cite{Luo2023}. They concluded that whether ChatGPT's impact is beneficial or harmful depends on its implementation and the balance stakeholders strike between its potential benefits, challenges, and risks. Their research flags several domains where ChatGPT could pose challenges, including ethical and social concerns, technical limitations, human oversight, and issues of accessibility and the digital divide.

Adding to the discourse, Murgia et al. explored the use of ChatGPT in education, with a particular emphasis on adapting online inquiries to meet different literacy levels \cite{10.1145/3563359.3596996}. Their findings indicate that while ChatGPT can adjust its responses to match a 4th-grade level, there remains room for improvement to achieve optimal readability.

Moreover, in self-guided learning environments where ChatGPT served as a digital mentor, it demonstrated potential benefits for adult learners in defining educational objectives, identifying resources, and tracking progress \cite{Lin2023}. However, scholars raised concerns about potential over-dependence on ChatGPT and the ambiguity of AI usage policies in different educational contexts \cite{Lin2023}.

In a societal analysis, Kristjan and Siibak drew attention to the tendency of tech companies to release tools like ChatGPT without fully considering their social implications \cite{Kikerpill2023}. They noted mixed reactions towards ChatGPT's introduction--some called for a ban in education, while others perceived it as a ``technological fix'' demanding a re-evaluation of educators' roles and teaching philosophies. They further warned against overlooking the commercial interests of OpenAI, considering its for-profit status \cite{Kikerpill2023}.

In conclusion, while AI and tools like ChatGPT appear to hold significant potential for enhancing learning experiences, concerns about accuracy, potential over-reliance, and unclarified AI usage policies persist. The current literature lacks a holistic exploration of AI-empowered environments, particularly from the perspectives of students, parents, and researchers. Therefore, further research is essential to fully comprehend human-AI interaction in educational settings. Our objective is to provide a comprehensive understanding of the topic by taking into account the perspectives of a wider spectrum of stakeholders involved.}

\section{Research Method} \label{sec:research_method}
\revis{The purpose of this study is to explore early adopters' experiences and thoughts about integrating ChatGPT into education by analyzing their social media content on various platforms, i.e., \textit{Twitter}, \textit{Reddit}, \textit{YouTube}, and \textit{LinkedIn}. Using social media content analysis, we can extract valuable themes and helpful insights from social media data on a larger scale \cite{Blandford2016}, ultimately leading to the identification of the primary applications (RQ1) and positive and negative perceptions of early adopters' use of ChatGPT in education (RQ2).

The remainder of this section provides detailed information on our data collection process, data description, analysis pipeline, and research ethics to provide a clear and comprehensive understanding of our qualitative study.

Given the advantages and challenges mentioned above of using ChatGPT-like chatbots in education, this study sought to identify the broader perspectives of early adopters (i.e., \textit{researchers}, \textit{educators}, \textit{learners}, \textit{parents}, and \textit{general users (unspecified roles)}), about the educational applications of ChatGPT to unravel its current deficiencies and strengths and discuss the potentials of future use. 

\subsection{Data Collection} \label{subsec:data_collection}
For this study, we leveraged the application programming interfaces (APIs) of Twitter \cite{twitterapi}, Reddit \cite{redditapi}, YouTube \cite{youtapi}, and  LinkedIn \cite{LinkedInapi} to gather data on the use of ChatGPT in education. Our data collection on LinkedIn was further enhanced by using the advanced (and legal) data scraping functions available on Helium (v.3) (similar to \cite{10.1145/3555124}).

With over 400 million monthly active users \cite{TwitterUsers}, Twitter provides an ideal platform for screening global users' opinions on ChatGPT and its educational applications. Reddit, with more than 50 million daily active users \cite{RedditUsers}, provides rich data through its various subreddits, including those discussing the use of ChatGPT in education. YouTube, with more than 2.5 billion monthly active users \cite{YouTubeUsers}, is the world's largest video-sharing platform and offers valuable long-form audio and visual data, such as seminars, lectures, and crash courses discussing the use of ChatGPT in education. Finally, LinkedIn, with more than 300 million monthly active users \cite{LinkedInUsers}, is the world's largest professional networking platform and provides extensive data on the professional use of ChatGPT in education. We selected these platforms because of their large user bases, rich data, accommodating terms of use, and potential to provide valuable insights into our research questions about the use of ChatGPT in education \cite{Sinha2021, Blandford2016}.

\textbf{Inclusion Criteria.} To ensure the relevance of the data collected, we used specific inclusion criteria. The content had to include the keywords ``ChatGPT'' AND (``Education'' OR ``Learn'' OR ``Teach'' OR ``Instruct'' OR ``Study'' OR ``School'' OR ``University''). We chose these keywords based on their prevalence of use in learning communities, as determined by Ubersuggest\footnote{https://neilpatel.com/ubersuggest/} (similar to \cite{10.1145/3430895.3460126}). Our analysis was conducted with the objective of minimizing the likelihood of overlooking crucial information---to the best of our ability---through the use of an array of keywords. It should be noted that all of the data used in our research were analysed in text format, but the scope of the data collection was restricted to content expressed in English. YouTube data were transcribed into textual format relying on the transcript feature that is accessible on the platform. In instances where a transcript was not provided by the content producer, we used the autogenerated transcript functionality offered by YouTube.

\textbf{Data sampling.} Because studying all social media data is impractical, we followed the advice of existing literature and collected random samples from highly viewed and liked social media posts, specifically those in the top 25\% (upper quartiles), on Twitter, Reddit, YouTube, and LinkedIn (refer to \cite{10.1145/3313831.3376768, 10.1145/3555124}). To accomplish this, we employed a stratified random sampling method, without personalization (see \cite{10.1145/3479556}), similar to the method used in a Computer Supported Cooperative Work (CSCW) study conducted by Fogliato et al. \cite{10.1145/3479572}. Our decision to use stratified random sampling was motivated by a desire to secure an impartial and equitable representation of early adopters' viewpoints, encompassing informed \textit{researchers}, \textit{educators}, \textit{learners}, \textit{parents}, and \textit{general users}. We established this by searching for the keywords presented in Table \ref{tab:earlyadoptersroles} within user profiles and manually annotating the data. The alternative keywords presented in Table \ref{tab:earlyadoptersroles} were recommended using a built-in function designed for similar tasks on the popular qualitative research application called \textit{Atlas.ti} \cite{Atlastiref}. Prior to stratified sampling, the majority of data only reflected the perspectives of researchers, educators, and general users.

\begin{table}[t!]
  \centering
  \caption{Alternative Keywords for Early Adopters (in Alphabetical Order)}
  \begin{tabularx}{\linewidth}{>{\raggedright\arraybackslash}X >{\raggedright\arraybackslash}X}
    \toprule
    Early Adopter Roles & Alternative Keywords \\
    \midrule
    Researcher & Academic, Analyst, Expert, Investigator, Professional, Scholar, Scientist \\
    Educator & Coach, Instructor, Mentor, Professor, Teacher, Trainer, Tutor \\
    Learner & Apprentice, Student, Trainee \\
    Parent & Father, Guardian, Mother \\
    General User (unspecified role) & N/A \\
    \bottomrule
  \end{tabularx} 
\label{tab:earlyadoptersroles}
\end{table}%

Using this sampling technique, we were able to obtain a diverse but manageable sample of social media posts. It is essential to note that choosing posts with a high number of views and likes is justifiable because they are more likely to contain content that resonates with a larger audience. This is crucial to understanding the dominant attitudes and conversations on these social media platforms~\cite{10.1145/3313831.3376768}.

\textbf{Exclusion Criteria.} To ensure the precision and authenticity of our analysis, we limited our focus to the experiences of early adopters with real and practical knowledge of ChatGPT in educational settings, which were manually verified by three co-authors through user-generated content. In compliance with the established literature, we intentionally excluded user comments, external links, and image-based content from our analysis to avoid an exponential increase in data volume. Our analysis solely incorporated comments from the Reddit platform because of its discussion-centric style of communication (see \cite{10.1145/3487553.3524202}). Examining comments and external links would have greatly complicated our efforts to authenticate and analyze user experiences, as these sources frequently lack nuanced details of their parent posts. Although we recognize that some of the excluded content may have provided valuable insights, we considered this exclusion necessary to maintain a more rigorous and systematic approach to data collection and analysis.
% Image-based qualitative research necessitates an entirely different coding methodology.
% There are also currently no AI-based annotation models publicly available to support such sort of task of verification. 

\textbf{Stopping the Data Collection.} As our data collection proceeded in parallel with our data analysis process, we stopped collecting data once we achieved thematic saturation, which is the point at which no new information could be extracted from our data. The decision for thematic saturation was unanimously agreed upon by all data coders involved in the research, comprising three co-authors. To prevent or reduce the chance of premature termination of data collection, we intentionally analyzed a minimum of 1,000 pieces of content from each social media platform, although thematic saturation occurred much earlier in most cases (i.e., on average = 825.69, SD = 145.66). In addition, we used 500 pieces of content to confirm the accuracy of our thematic saturation as a final measurement. 

\subsection{Data Description} \label{subsec:data_description}
The data analyzed in this work cover the time period between December 1st, 2022, and April 22nd, 2023, encompassing the five months following the official public release of ChatGPT. Our study examines 6,000 text
samples totaling roughly 150,000 words to determine how people commonly used ChatGPT in various learning contexts. Compared to similar prominent HCI studies, the size of our dataset seems reasonable to provide a broad view of early users' opinions (e.g., see \cite{10.1145/3313831.3376768, 10.1145/3290605.3300881, 10.1145/3430895.3460126}).

More specifically, our dataset includes 1,500 randomly selected posts from each of the four social media platforms (see Section \ref{subsec:data_collection}), with equal numbers from five user groups in education: researchers (300), educators (300), students (300), parents (300), and general users (300). We used 1,000 posts per platform for the thematic analysis, reserving the remaining 500 (100 from each user group) to confirm the final saturation of the themes. Figure \ref{fig:snfigure} displays the distribution of user discussions concerning the utility of ChatGPT across diverse educational settings.

\begin{figure*}[]
    \centering    \includegraphics[width=0.6\textwidth,keepaspectratio]{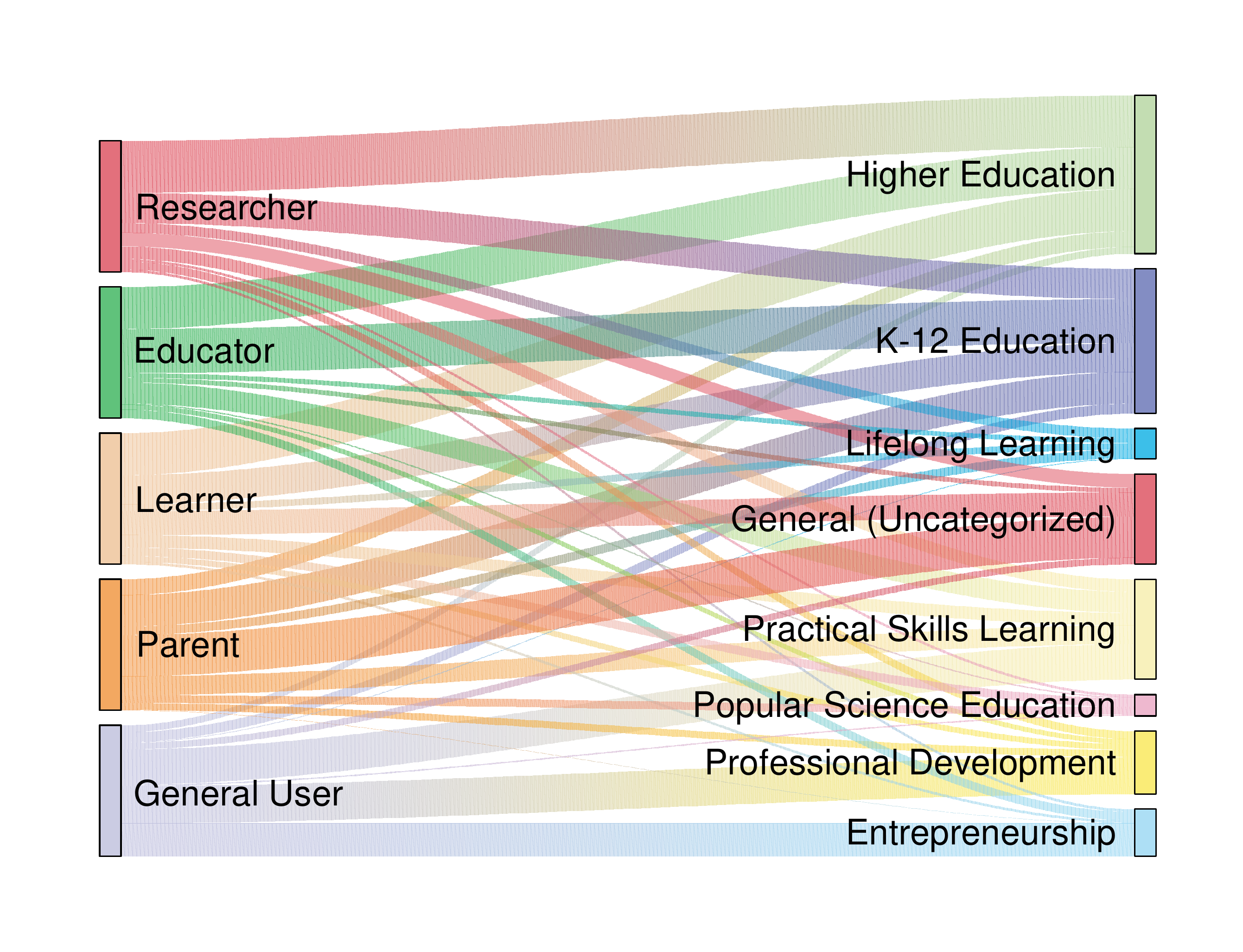}
    \vspace*{-5mm}
    \caption{Distribution of user discussions concerning the utility of ChatGPT across diverse educational settings. ChatGPT is primarily discussed in the context of higher education, K-12 education, and practical skills learning.} \label{fig:snfigure}
\end{figure*}

According to our analysis, most posts (55.27\%) came from users in North America (34.23\%) and Northern Europe (21.04\%). The rest came from: Oceania (4.10\%), Asia (2.88\%), Africa (1.17\%), and elsewhere (0.98\%). It should be noted that for 35.60\% of the data, the user's location is either not provided or disallowed for collection by the social media platform. 

The use of ChatGPT in education platforms is a more frequently discussed topic in wealthier, more developed northern countries. This observation, however, could also be attributed to regional language differences as well as the dominance of other local social media platforms in other areas. For example, WeChat, Sina Weibo, and Zhihu are more popular social media platforms in China \cite{10.1145/3411763.3451717, 10.1145/3411764.3445436, 10.1145/3500931.3501025}. Furthermore, it should be noted that as of April 25, 2023, ChatGPT is not accessible or has stopped providing services in countries such as China, Cuba, Iran, Italy, Russia, Syria, Uzbekistan, and Venezuela \cite{chatgptavail}.
\subsection{Data Analysis} \label{subsec:analysis_pipeline}
In this study, we used a qualitative research approach, specifically the inductive coding method, to analyze the collected social media data. We decided to conduct a qualitative study because this type of research is best suited to capture the complex and nuanced aspects of users' reactions and responses to new technologies, trends, and phenomena (see \cite{Blandford2016, 10.1145/3555124}). The inductive method is particularly suitable for our investigation because of the apparent lack of prior empirical research on user experience related to the use of ChatGPT in education. Furthermore, the inductive coding approach enables us to develop themes from the data without preconceived notions or prior assumptions \cite{Blandford2016, 10.1145/3555124}. 

We used multiple digital tools to facilitate our coding process. Microsoft Excel helped organize and manage our data. Atlas.ti \footnote{https://atlasti.com/}, a qualitative data analysis software, allowed for collaborative human-AI sentiment analysis\footnote{The default categories are \textit{positive}, \textit{negative}, and \textit{neutral}}, keyword searching, data coding, and report writing. Jamboard \footnote{https://jamboard.google.com/}, a virtual whiteboard, allowed brainstorming and visualization of emerging themes. The combination of these tools enabled a thorough and rigorous analysis of our data. Figure \ref{fig:alltools} shows the snapshots of these tools used in our research.

\begin{figure*}[t!]
    \centering
    \subfigure[Microsoft Excel]{\includegraphics[width=0.45\textwidth]{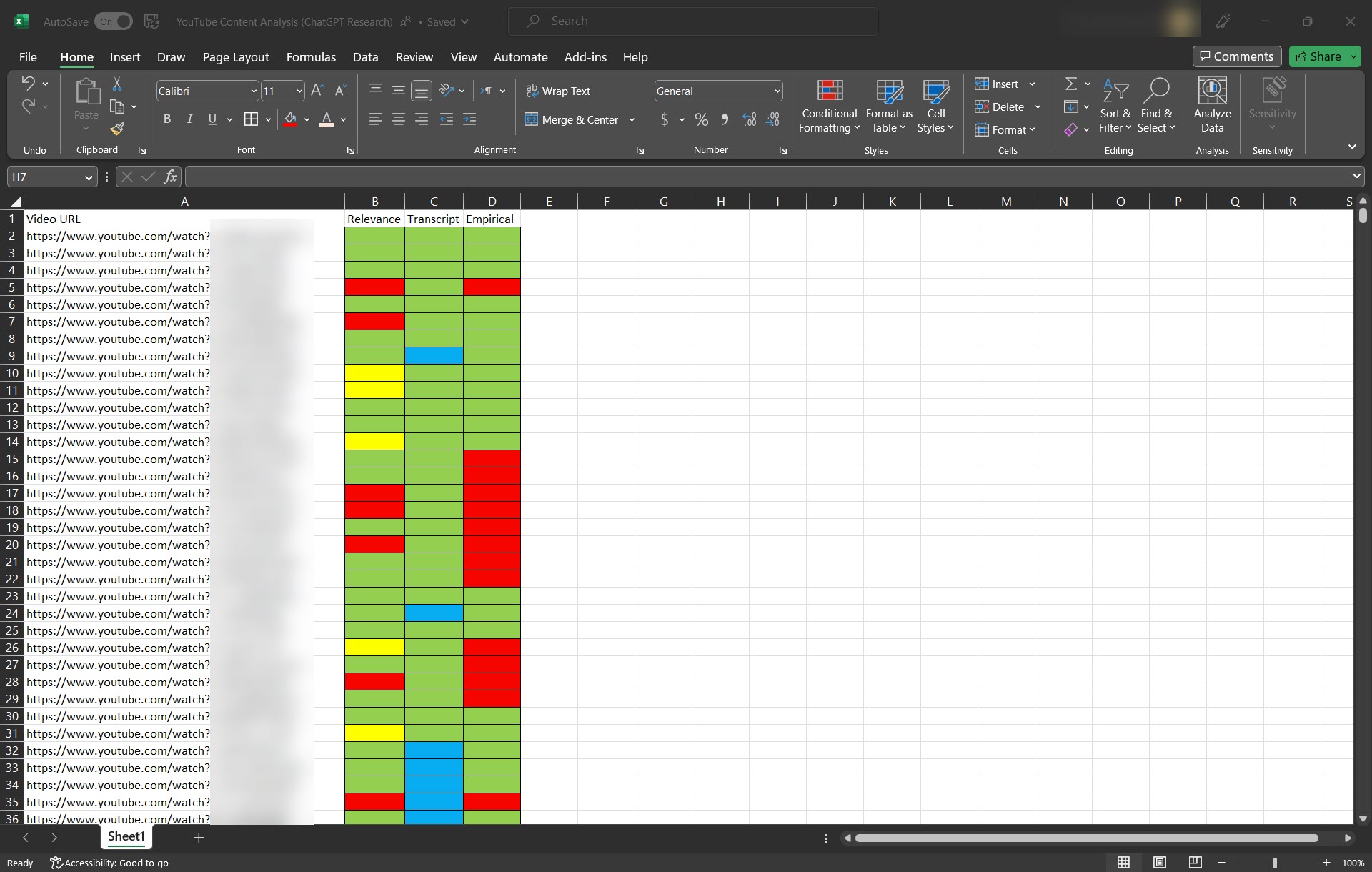}}
    \subfigure[Coding with Atlas.ti]{\includegraphics[width=0.45\textwidth]{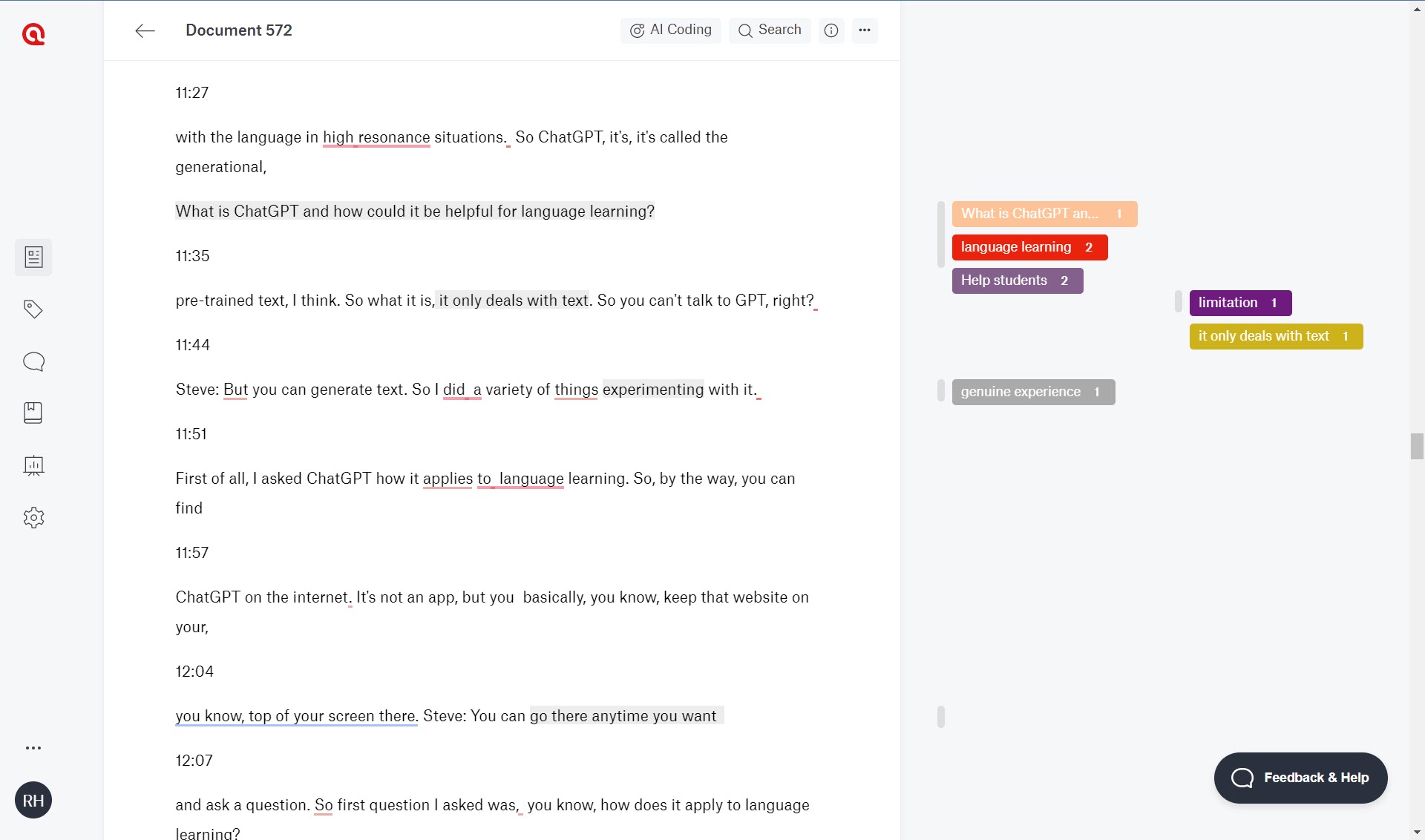}}
    \subfigure[Jameboard Discussion Board]{\includegraphics[width=0.45\textwidth]{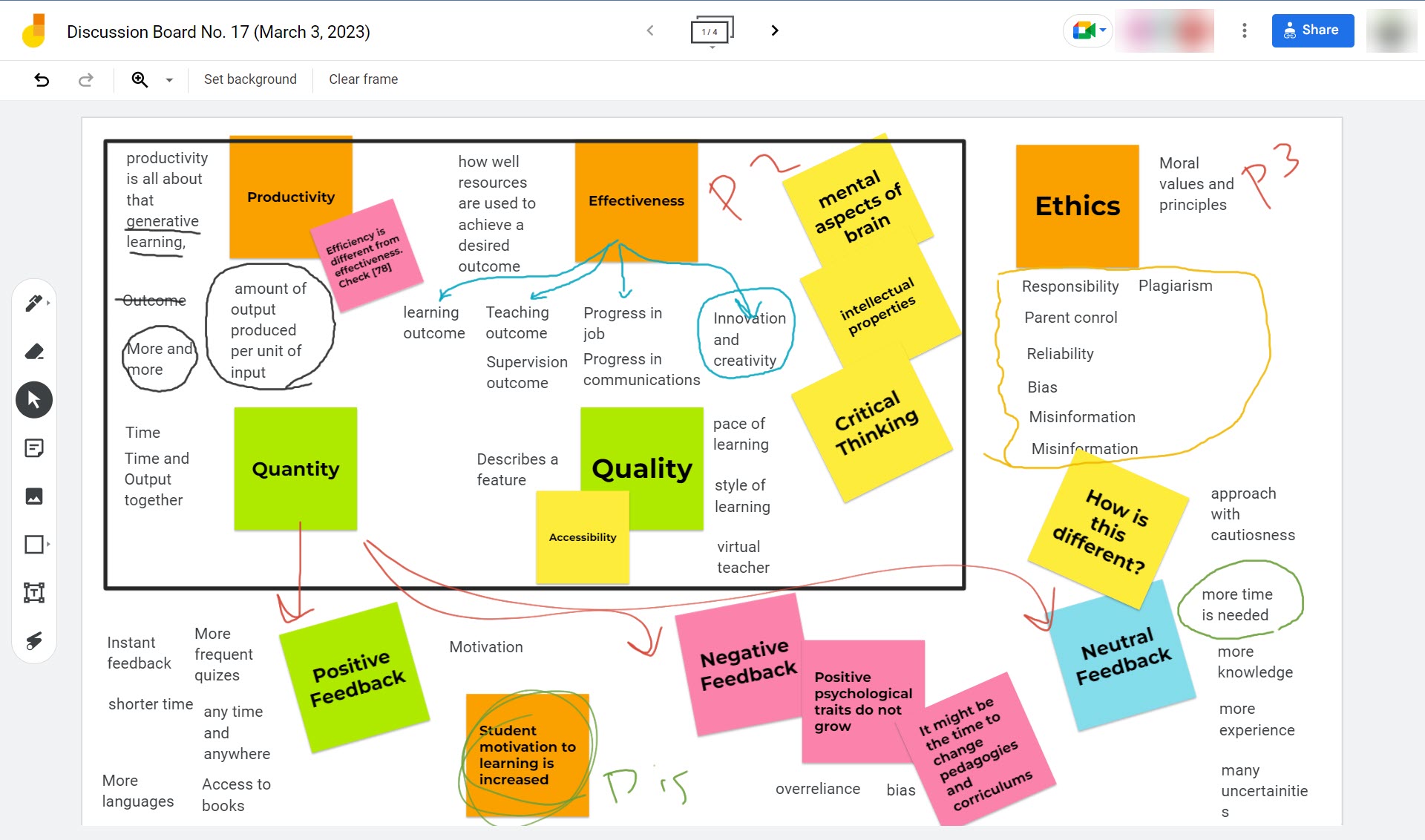}}
    \caption{Snapshots of tools used during the coding process}
    \label{fig:alltools}
\end{figure*}

\textbf{Analysis Pipeline.} To identify the overarching themes in our data, we employed Braun and Clarke's (2006) \cite{Braun2006} six-phase framework (similar to \cite{Jeon2023}). This involved becoming familiar with the data, creating initial codes, searching for themes, reviewing them, consolidating titles and definitions, and finally reporting the themes. To obtain a more comprehensive understanding of people's opinions and minimize the chances of unintentional biases as much as possible, data from all four social media platforms were analyzed concurrently (similar to \cite{10.1145/3430895.3460126}).

Three authors (henceforth: coders) conducted the thematic analysis through an iterative coding procedure collaboratively. The third coder was brought in to help resolve any potential disagreements between the first two coders. 

In the initial phase of coding, each of the three coders independently analyzed the data and generated an initial set of codes. The coders then met remotely via video conference using the Zoom application to review the codes identified individually, discuss any disagreements in coding, and arrive at a consensus on the initial set of codes. From this initial meeting, we compiled a codebook containing a summary of the codes and examples of data extracts (quotes) for each code.

We went through the independent coding process, followed by group discussion and consensus, a total of five times over two months. With each iteration, we refined the codes by collapsing some codes and separating others into distinct concepts. We also identified emerging themes throughout the codes. After the fifth round of coding and discussion, we arrived at a final codebook that contained the most important information in the data and answered our research questions.

As a final step in the thematic analysis process, we used the Fleiss' Kappa measure to estimate and report the level of agreement among all coders. We calculated the Kappa value using the reliability analysis tool in the \textit{IBM SPSS Statistics}\footnote{https://www.ibm.com/products/spss-statistics} application. Our study produced a Kappa score of $\kappa = 0.87$, indicating almost unanimous agreement among our team members on the final themes identified \cite{10.1145/3555124}.}
\subsection{Research Ethics} \label{subsec:research_ethics}
In this study, we adhered to established ethical guidelines for collecting and analyzing data from social media platforms \cite{GDPRref, AoIRref, 10.1145/3555124}. Before data collection, we obtained ethical approval from our institution's research ethics board to conduct this study. Furthermore, we obtained proper Application Programming Interface (API) keys from the platforms to legally access their data without overburdening their systems. We meticulously followed each platform's terms of service to ensure compliance with platform policies and user privacy. Furthermore, we only collected publicly available data and anonymized them by removing any personally identifying information. Furthermore, it should be noted that all coders in this study have received certified training in human subjects protection regulations, policies, and ethical issues.
\section{Findings} \label{sec:findings}
The study's findings are divided into two sections based on the research questions posed: (1) uncovering ChatGPT's most frequent applications in education, and (2) exploring different perspectives on ChatGPT's utility in education. The main themes within each
section are presented as bullet points. Quotes from the content analysis are indicated in italics, and the sources of different quotes and the positions (roles) of each commenter are reported while ensuring anonymity. Moreover, the percentage values indicate the frequency with which a topic appears during the content analysis.

\subsection{RQ1: Exploring the Multifaceted Uses of ChatGPT in Education}
\begin{figure*}[]
    \centering
    \subfigure[Higher Education]{\includegraphics[width=0.24\textwidth]{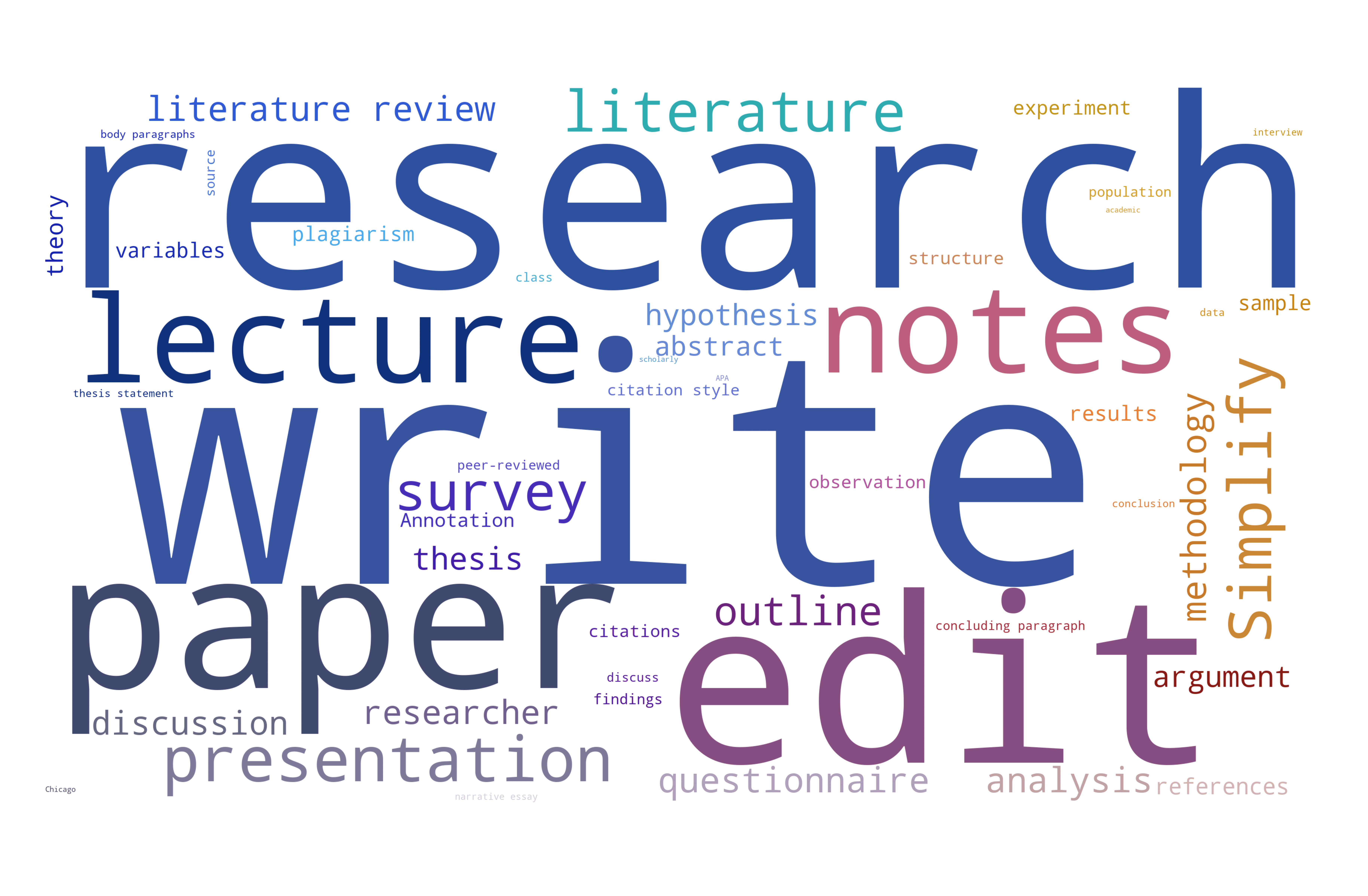}}
    \subfigure[K-12 Education]{\includegraphics[width=0.24\textwidth]{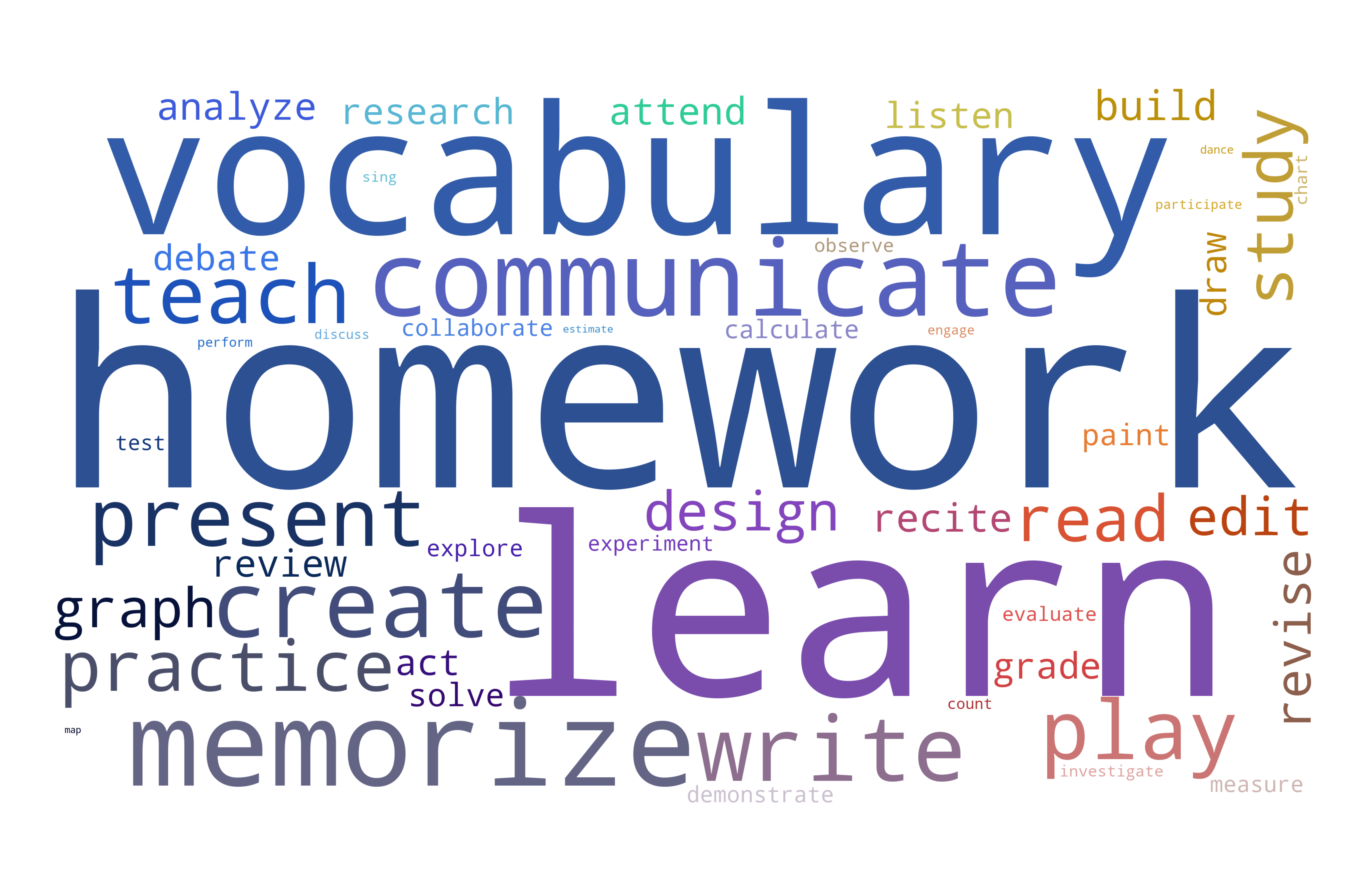}}
    \subfigure[Practical Skills Learning]{\includegraphics[width=0.24\textwidth]{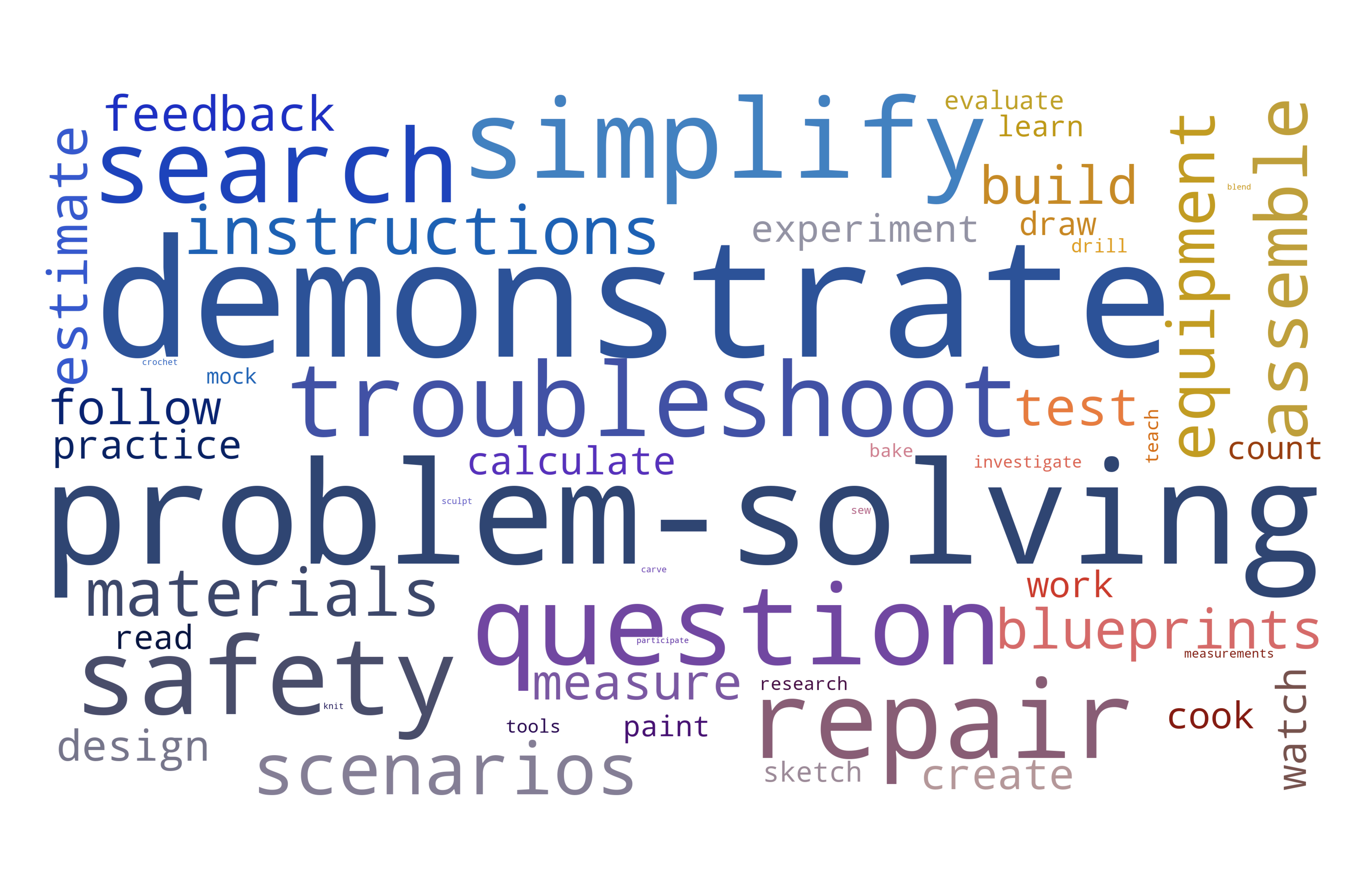}}
    \subfigure[Professional Development]{\includegraphics[width=0.24\textwidth]{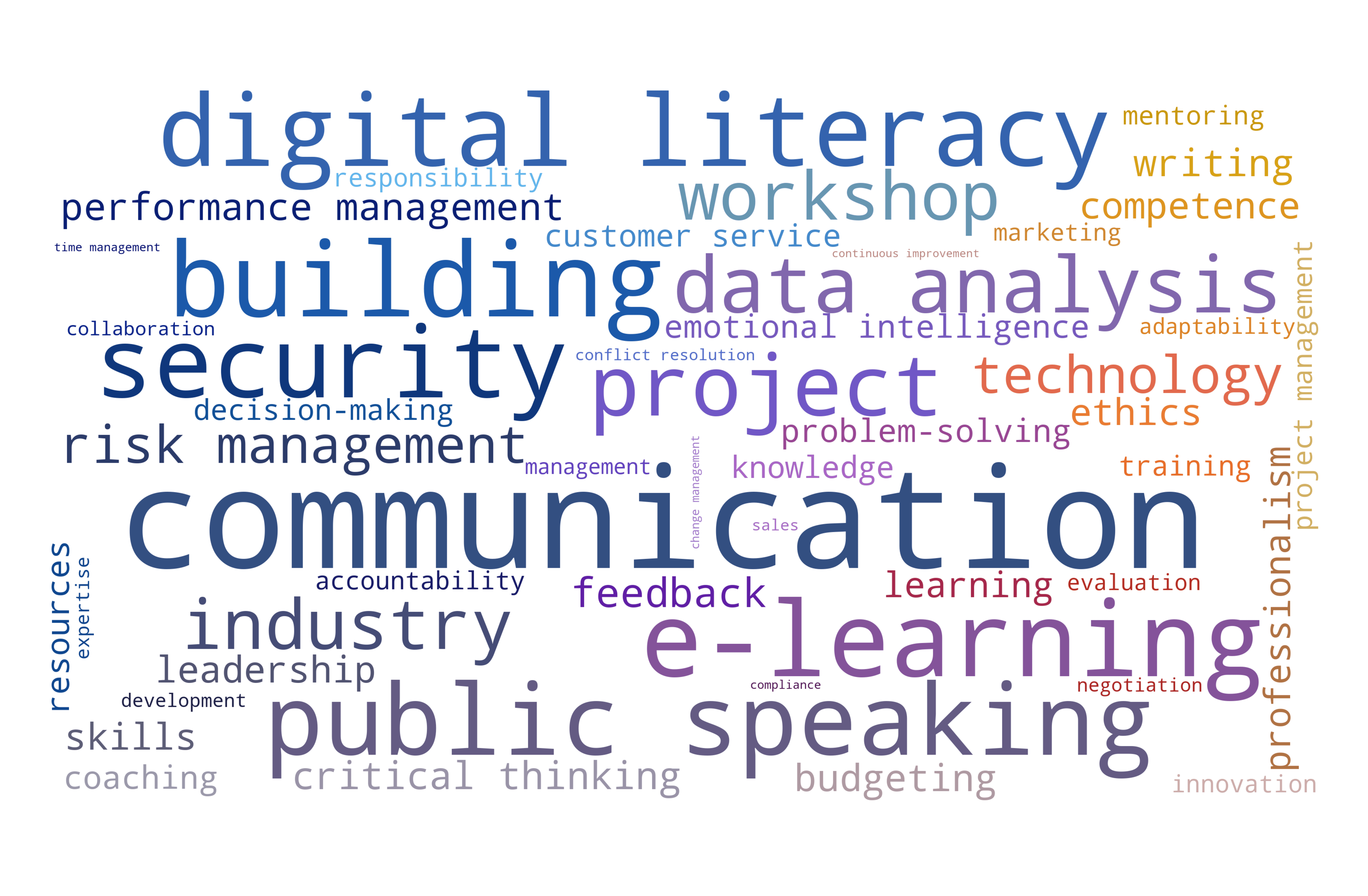}}\\
    \subfigure[Entrepreneurship]{\includegraphics[width=0.24\textwidth]{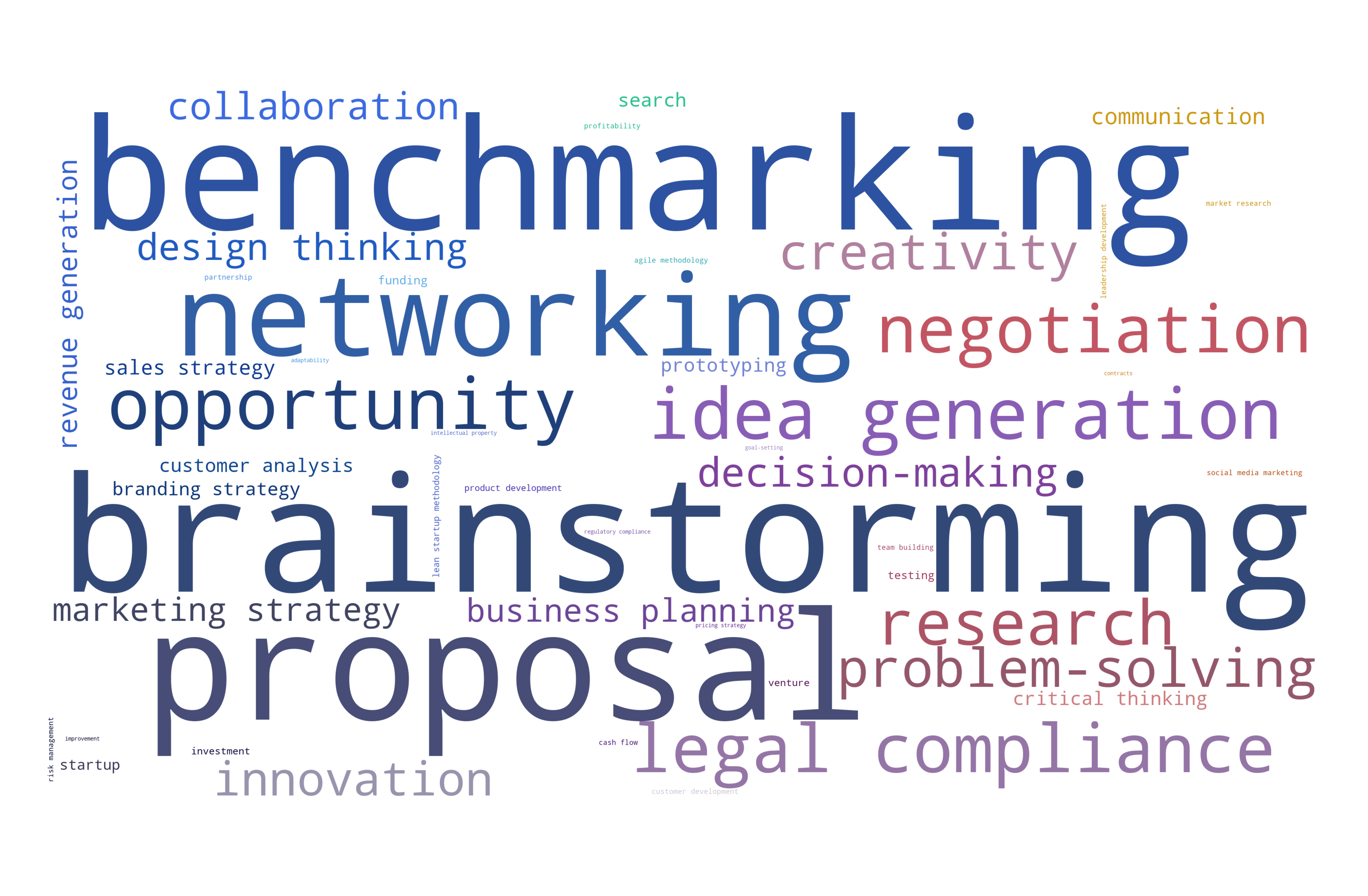}}
    \subfigure[Lifelong Learning]{\includegraphics[width=0.24\textwidth]{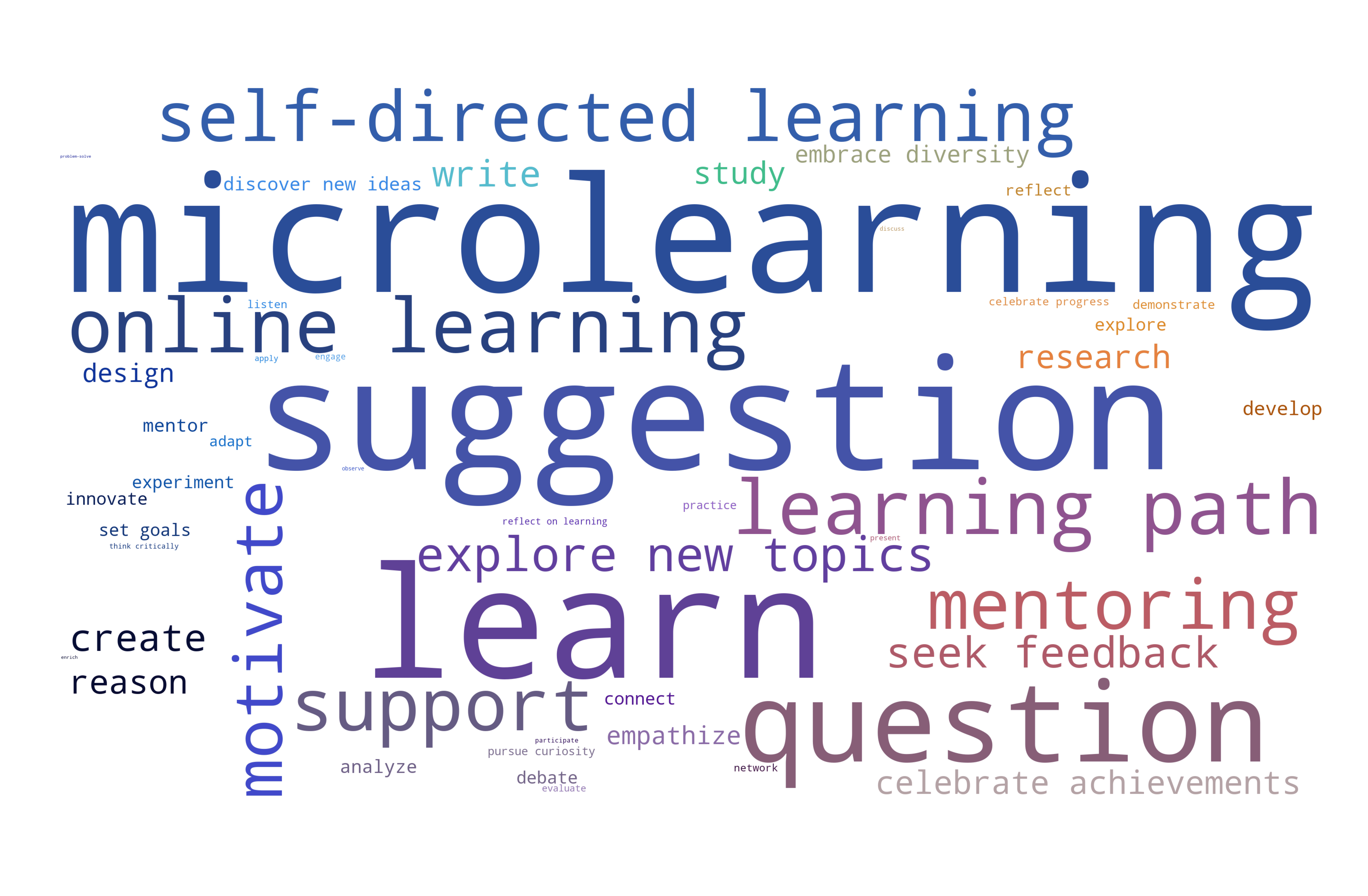}}
    \subfigure[Popular Science Education]{\includegraphics[width=0.24\textwidth]{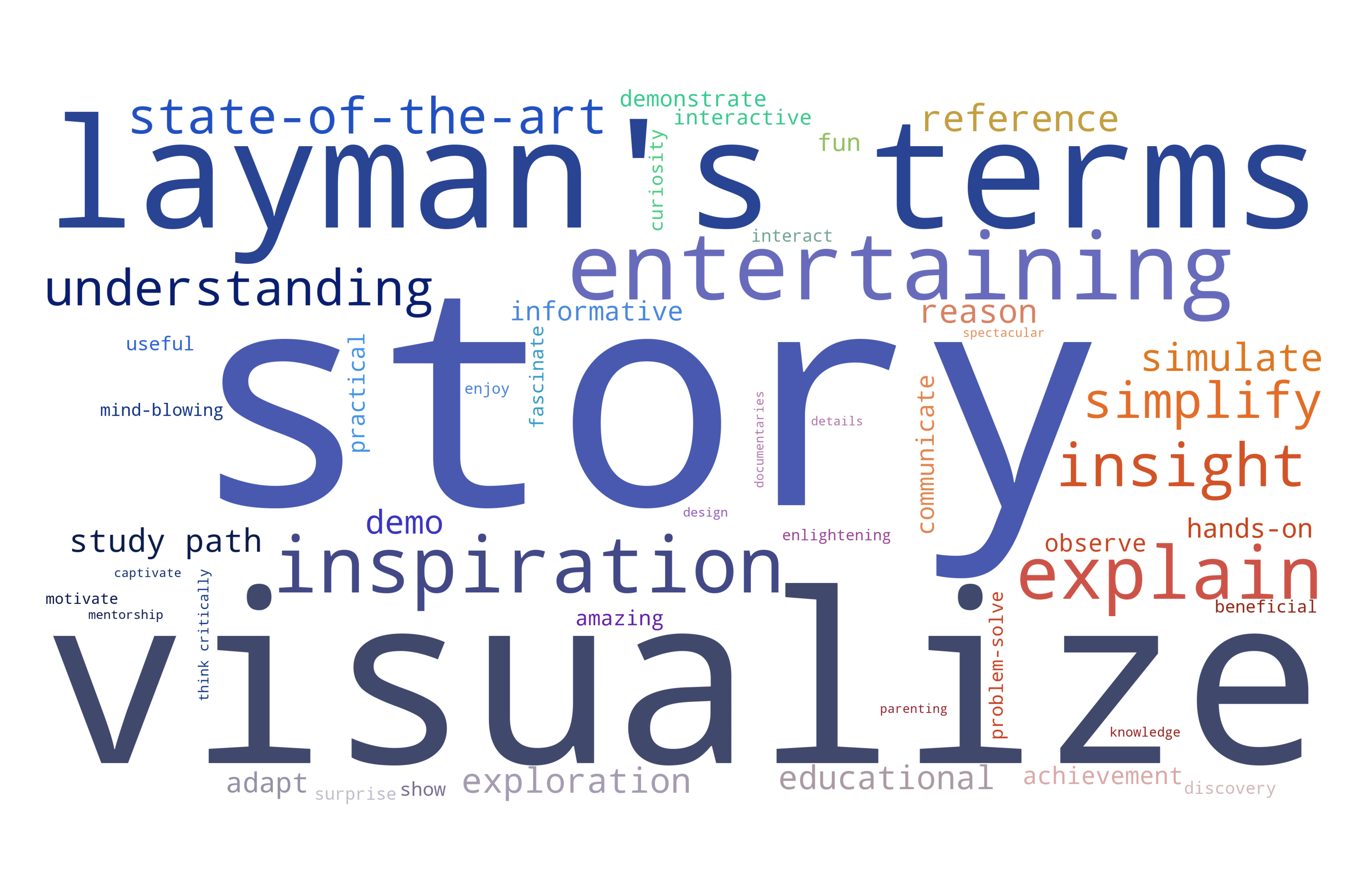}}
    \subfigure[General Learning Domain (Uncategorized)]{\includegraphics[width=0.24\textwidth]{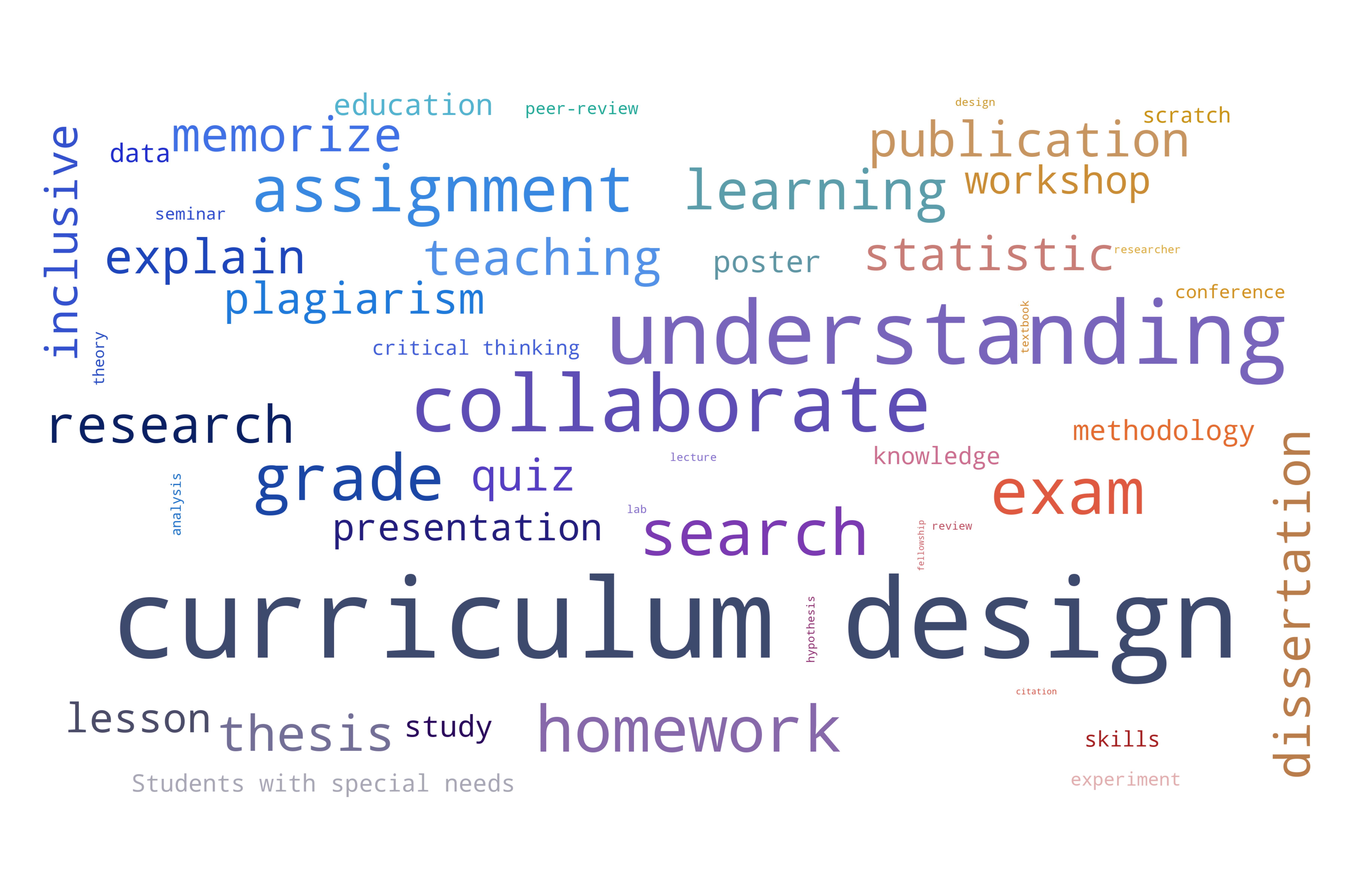}}
    \caption{The word clouds in this figure depict the top fifty applications of ChatGPT within various learning domains, sourced from social media platforms.}\label{wcfigs}
\end{figure*}

Our research reveals that ChatGPT is most frequently used in three distinct educational settings: higher education (24.18\%), K-12 education (22.09\%), and skills training (15.28\%). Other areas where ChatGPT has been used include professional development (9.61\%), entrepreneurship education (7.18\%), lifelong learning (4.64\%), and popular science education (3.25\%). Furthermore, 13.77\% of the instances are related to other learning domains that are not explicitly categorized. To visually represent the widespread use of ChatGPT in education, Figure \ref{wcfigs} displays the word clouds from social media-generated content, showcasing the top 50 applications of ChatGPT within each learning domain. In the following sections, we explore each category in more detail.

\subsubsection{Higher Education (24.18\%)} 
In the context of higher education, which refers to tertiary education following secondary education, typically in a college or university setting, ChatGPT has shown significant potential in helping researchers, educators, and students. 

$\bullet$ \textit{Content Creation and Editing.} The most prevalent use of ChatGPT in higher education is content creation and editing, accounting for 78.11\% of the reported applications. Educators, students, and researchers have used ChatGPT to write and proofread research papers, prepare lecture notes, create class presentation slides, and compose literature reviews. One professor shared her experience on YouTube, stating: 

``\textit{Simply enter the article or book chapter that you want to present in your next meeting or class and then presto! You have a nice outline for your slides, and the content can be edited however and as how many times as you like!}''

$\bullet$ \textit{Content Summarization.} Furthermore, ChatGPT has proven helpful in summarizing academic papers, as shared by a Ph.D. student on LinkedIn:

``\textit{ChatGPT is awesome for getting a quick rundown of authors' works. I'm finding it super helpful lately, and it could be a great way to sift through articles for a lit review and figure out what's [relevant] and worth reading and what's not.}''

 $\bullet$ \textit{Collaborative Data Analysis.} The integration of ChatGPT in higher education has not only streamlined various processes but also inspired new methods of research and data analysis. A distinguished professor of social science and humanities has proclaimed on their YouTube channel, ``\textit{ChatGPT is the beginning of a new era of big data exploration.}'' ChatGPT can serve as an AI-based ``\textit{research assistant for text-based data analysis, annotation, and sentiment analysis.}'' A helpful prompt this professor employs to improve their thematic analysis is asking ChatGPT, ``\textit{What are other angles this [issue, topic, or quote] could be analyzed from?}''

$\bullet$ \textit{Student Evaluation.} Teaching Assistants have found value in using ChatGPT for student evaluation. One TA shared on Twitter:

``\textit{ChatGPT saved me a lot of time and effort while grading 80 C++ homework submissions! I requested ChatGPT's assistance in creating a grading [rubric], and it was fantastic. I was able to review all the submissions, test them more [effectively], and even detect some instances of plagiarism.}''

$\bullet$ \textit{Peer-Review Processes.} Finally, the peer-review process is another area where ChatGPT has made an impact. A professor of HCI shared his experience on LinkedIn:

``\textit{Time is money ... [So] these days, before agreeing to review [a paper], I use ChatGPT to first get an overview of [the work] and assess its overall quality. A low ChatGPT score [ChatGPT-based evaluation] will automatically disqualify a paper from my review [list]. ChatGPT does a great job of summing up the key takeaways and contributions of the research. ... I highly recommend it to busy colleagues and peer reviewers!}''

\begin{figure*}[]
    \centering
    \subfigure[Higher Education]{\includegraphics[width=0.24\textwidth]{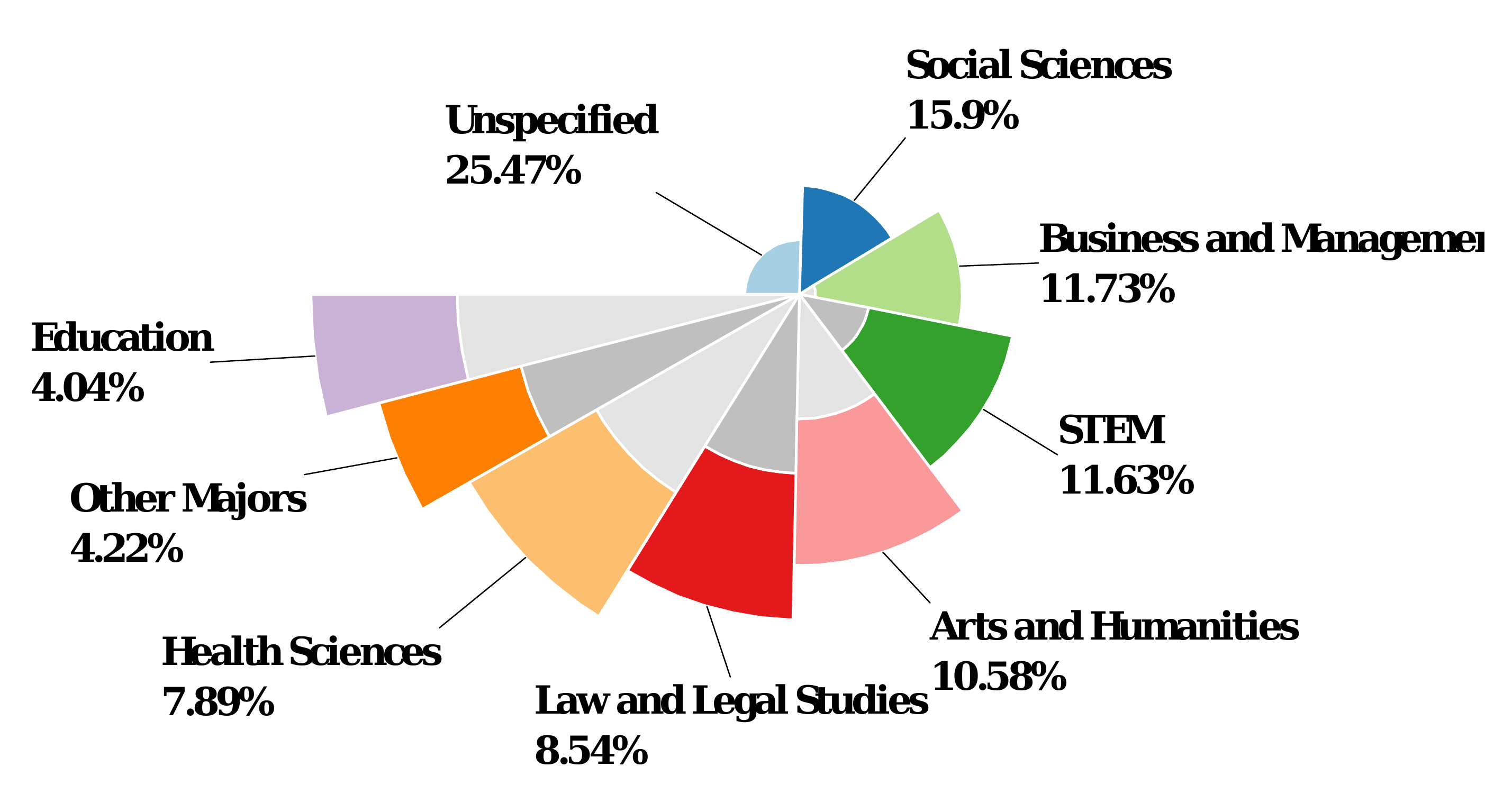}}
    \subfigure[K-12 Education]{\includegraphics[width=0.24\textwidth]{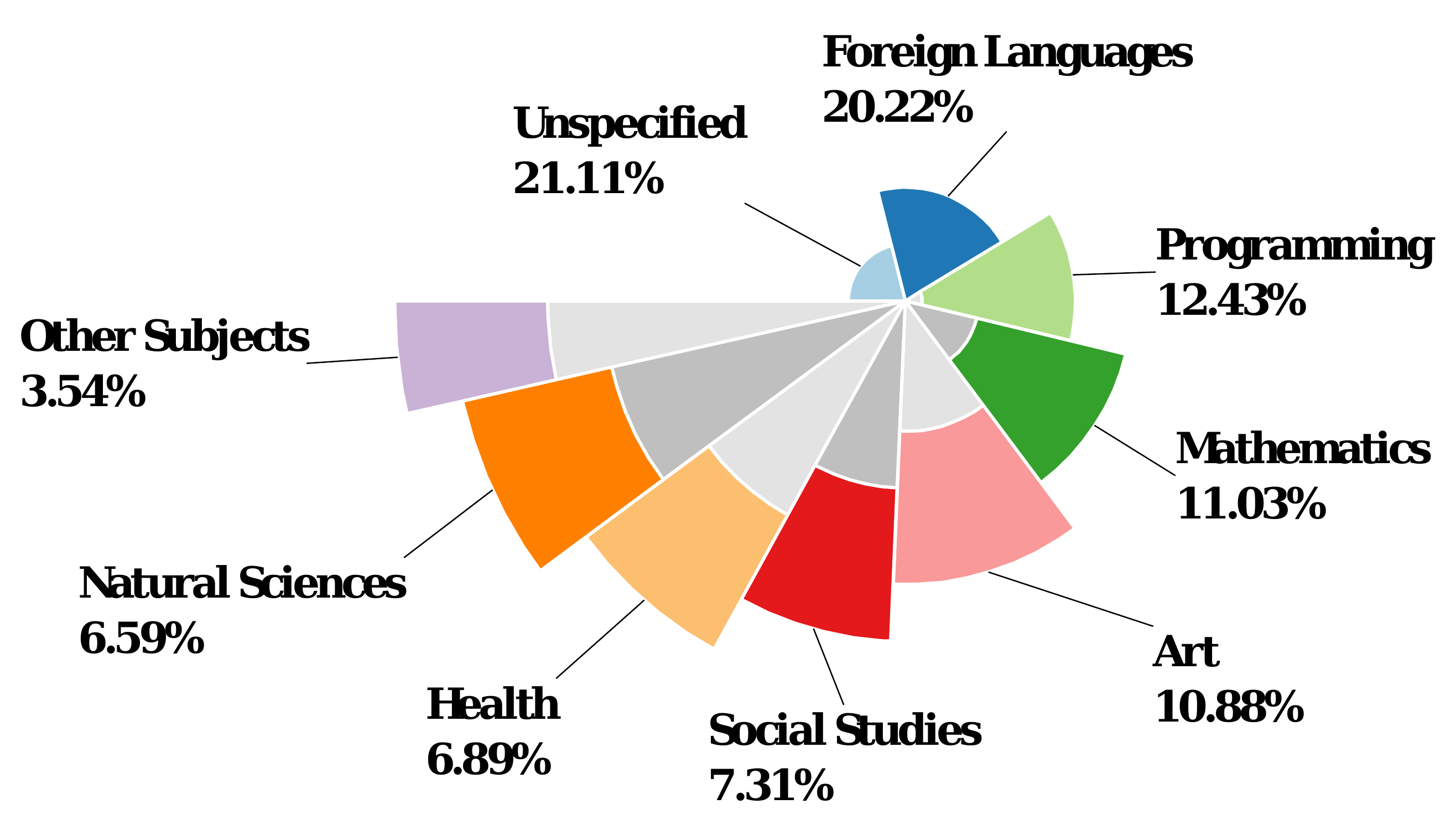}}
    \subfigure[Practical Skills Learning]{\includegraphics[width=0.24\textwidth]{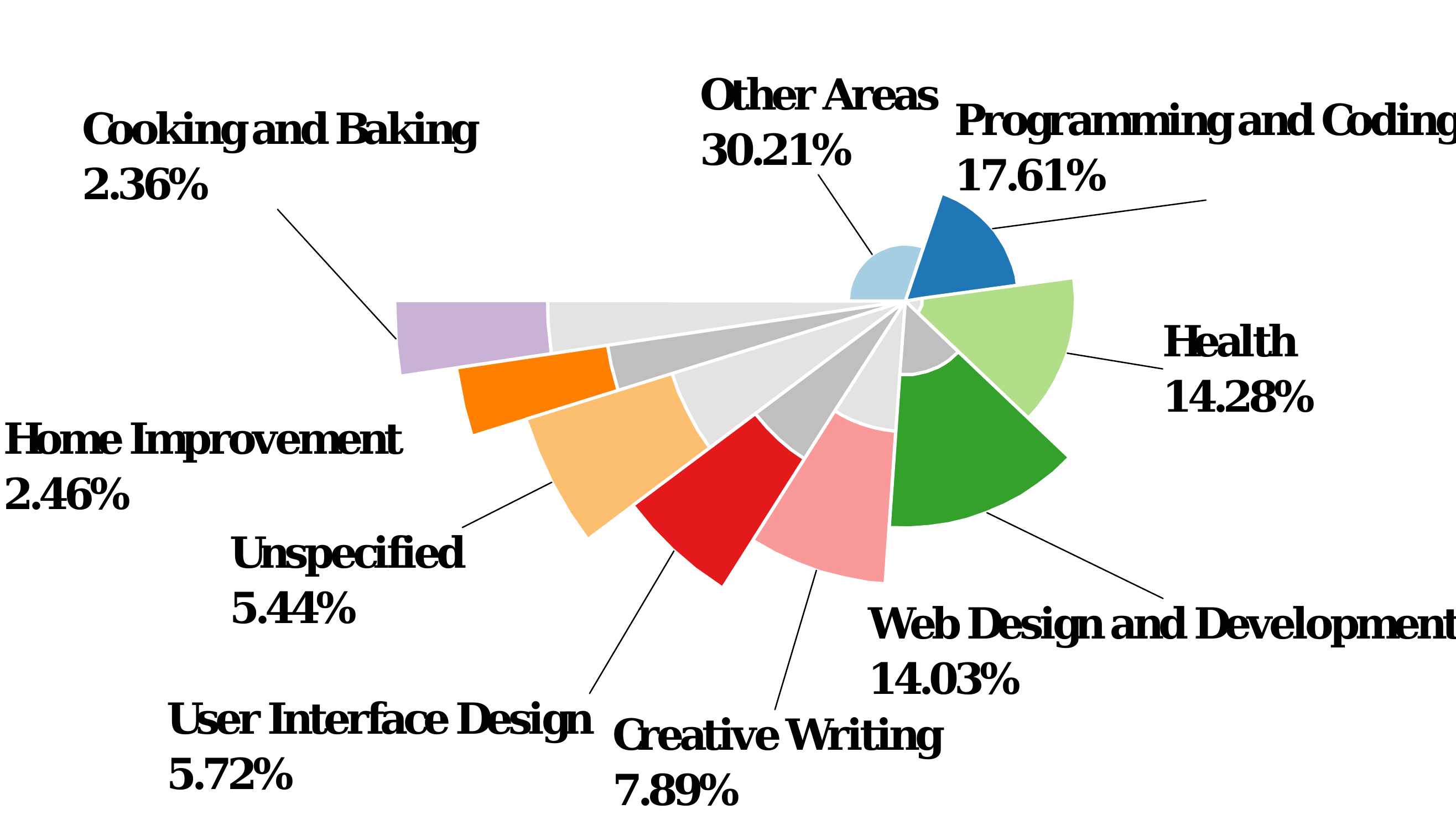}}
    \subfigure[Professional Development]{\includegraphics[width=0.24\textwidth]{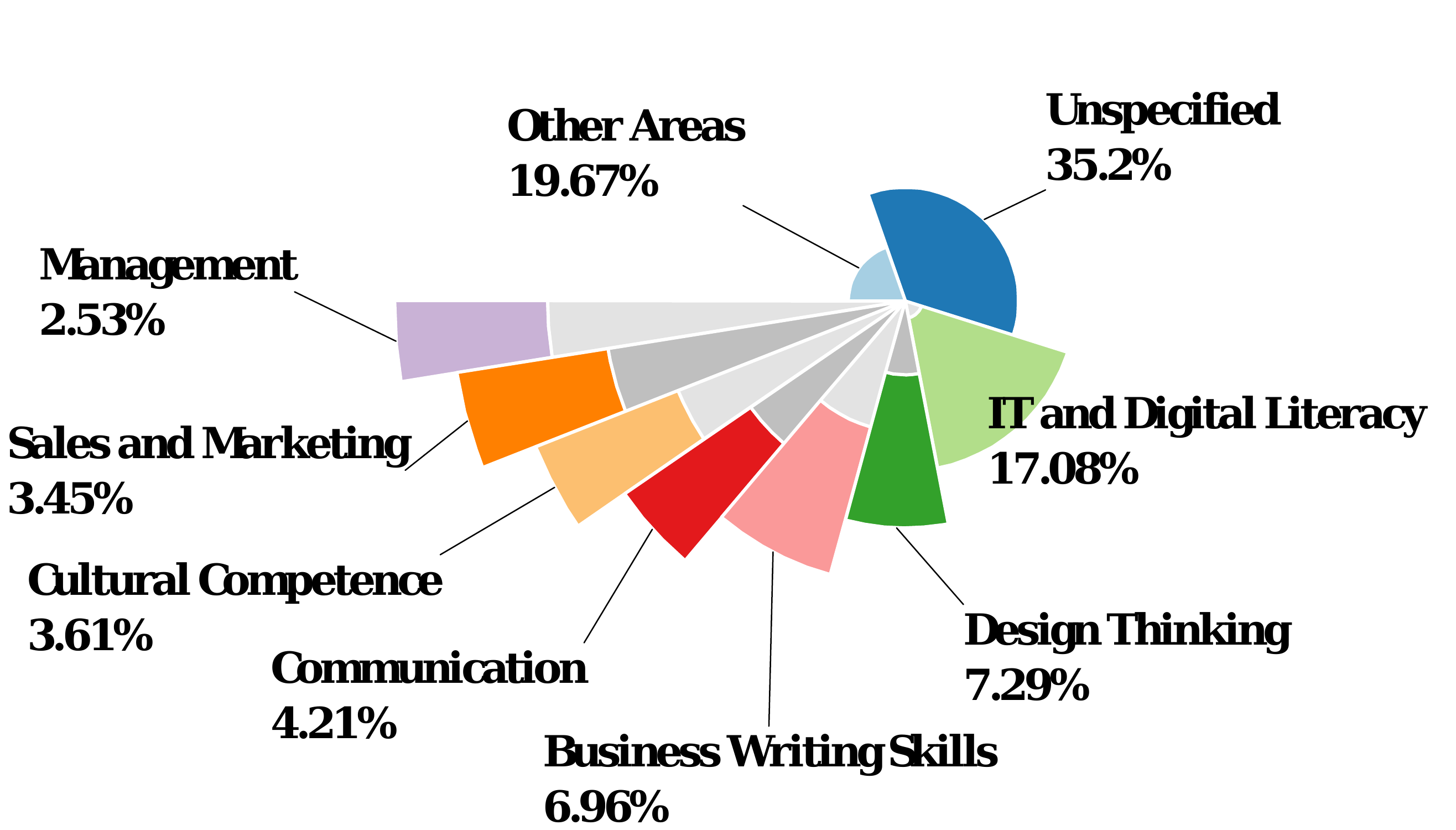}}\\
    \subfigure[Entrepreneurship]{\includegraphics[width=0.24\textwidth]{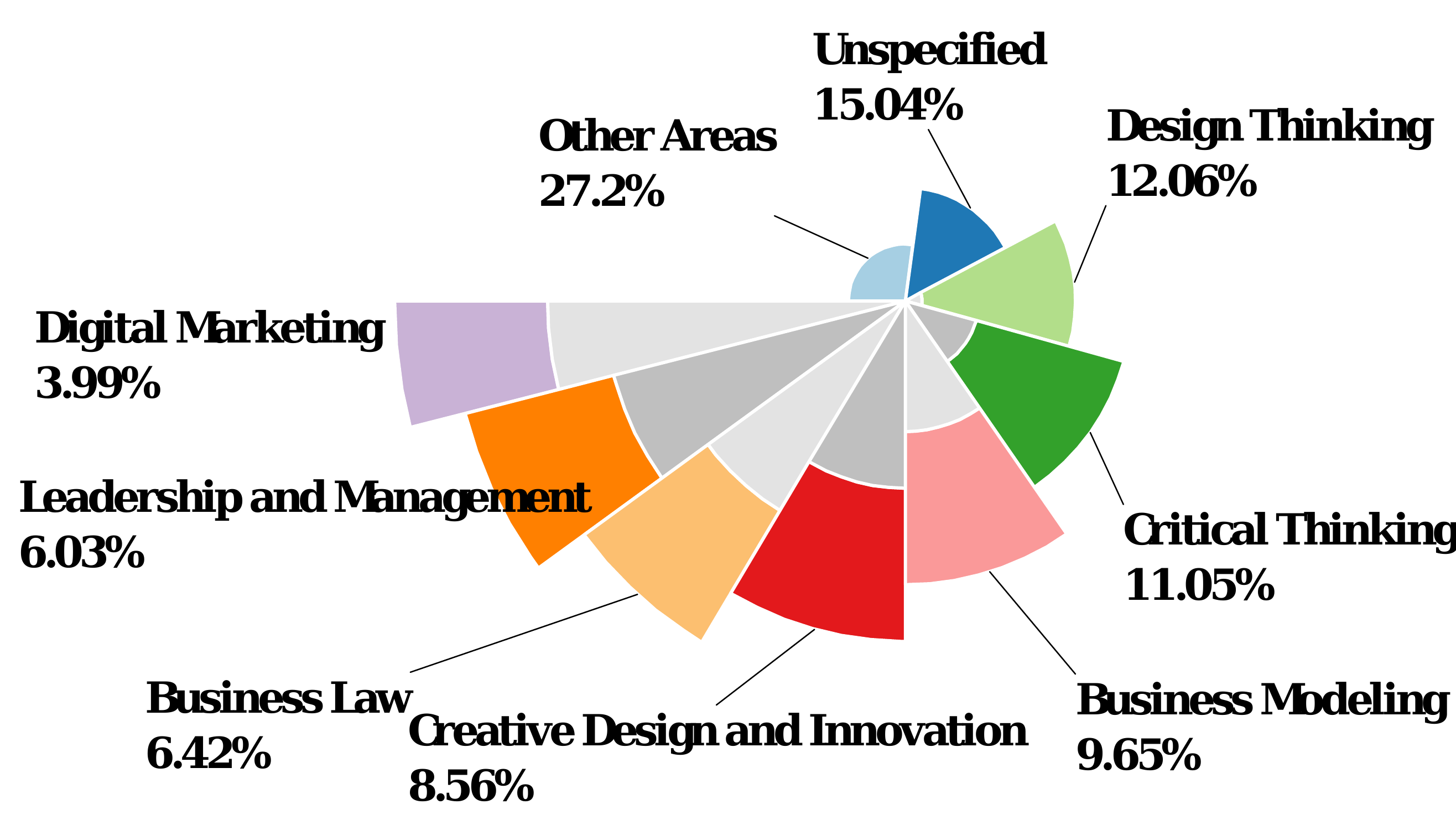}}
    \subfigure[Lifelong Learning]{\includegraphics[width=0.24\textwidth]{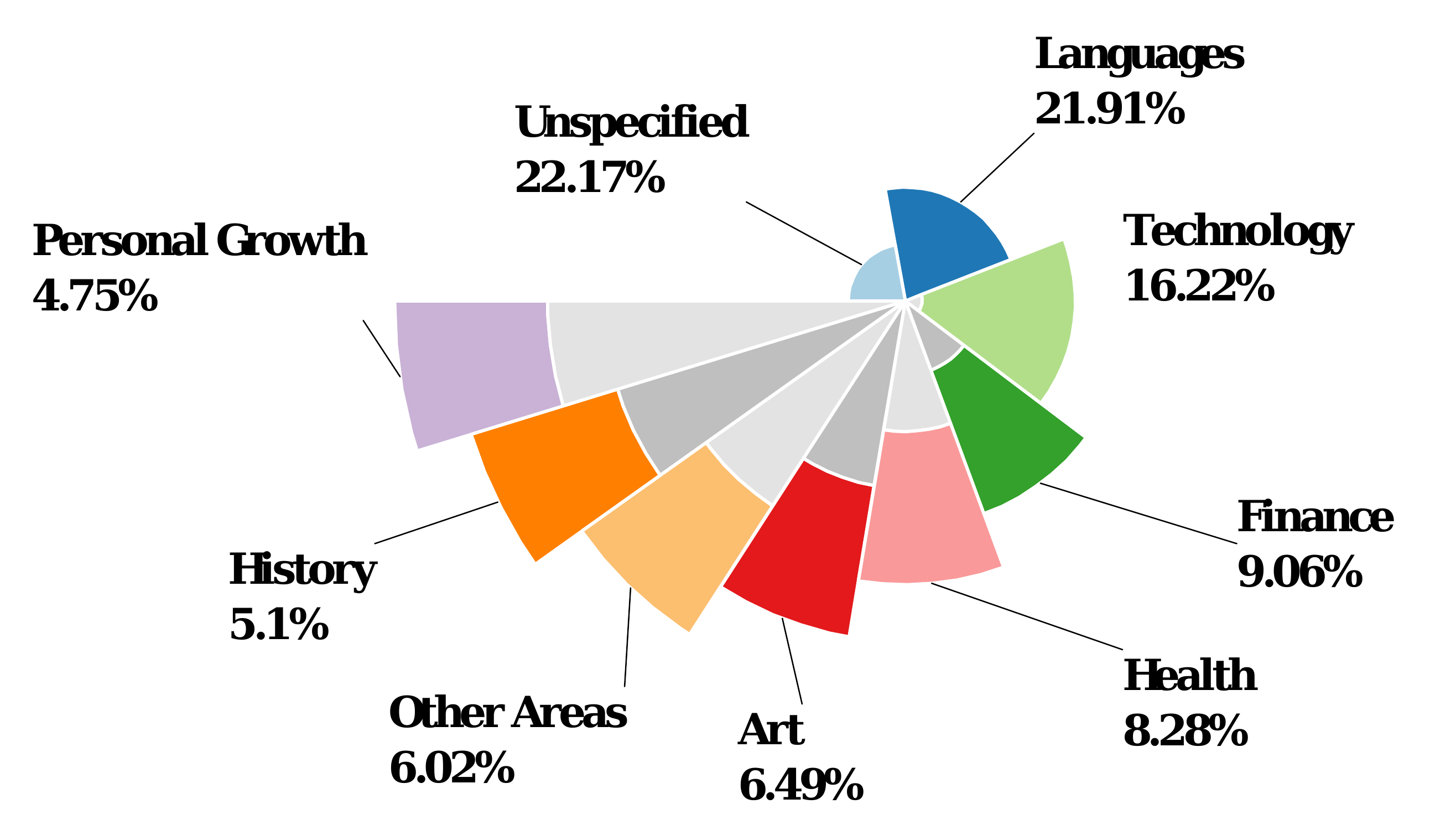}}
    \subfigure[Popular Science Education]{\includegraphics[width=0.24\textwidth]{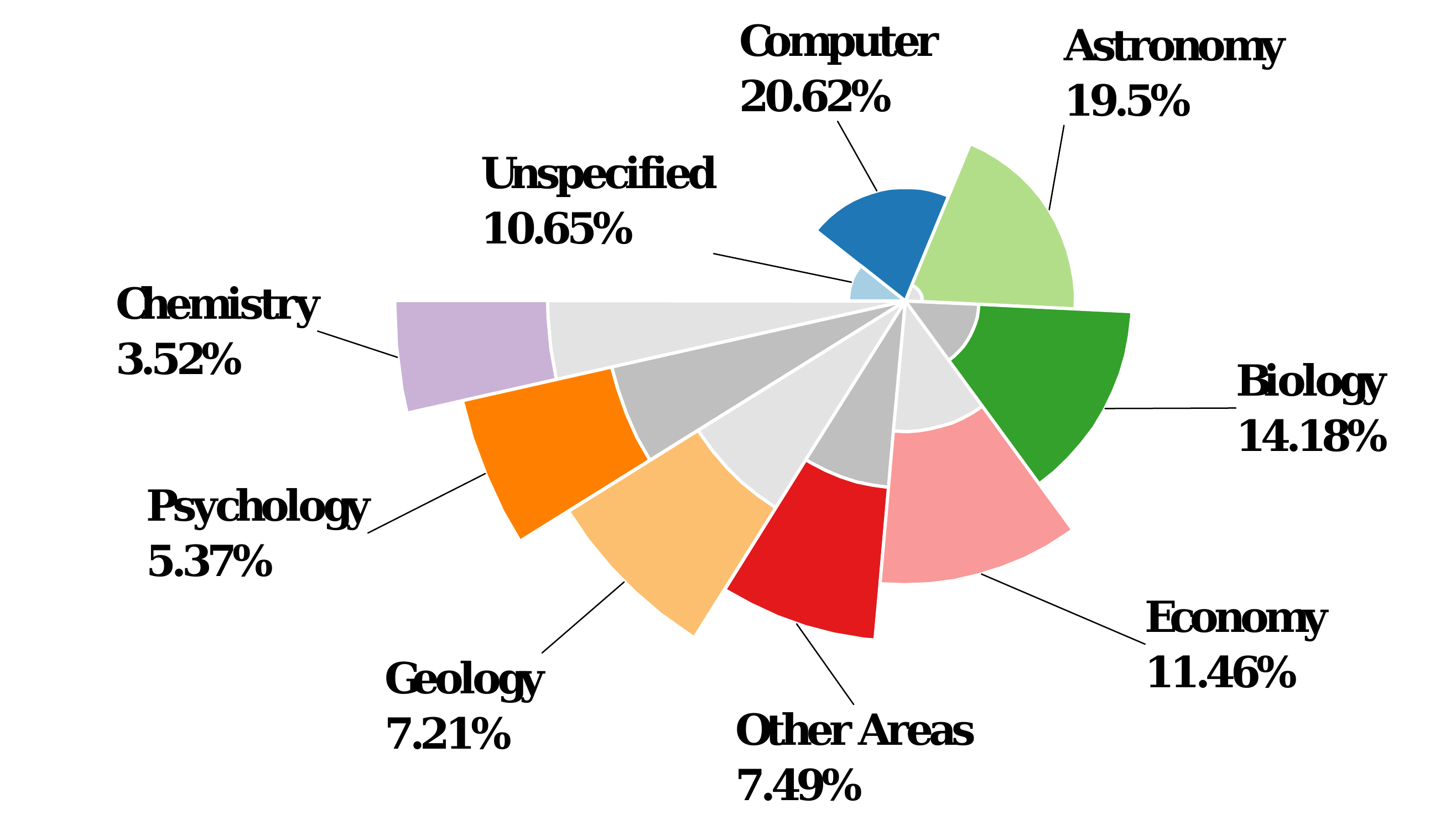}}
    \caption{Distribution of ChatGPT usage across different disciplines, study subjects, and learning areas}
    \label{fig:gptusageacross}
\end{figure*}

\textbf{Early Adopter Majors.} According to our findings, the three main subjects in which ChatGPT has been most frequently utilized by early adopters in higher education are \textit{Social Sciences} (15.9\%), \textit{Business \& Management} (11.73\%), and \textit{STEM Education} (Science, Technology, Engineering, and Mathematics) (11.63\%). Figure \ref{fig:gptusageacross} provides a more comprehensive overview of the most popular fields of study in which ChatGPT is reportedly used the most.

\subsubsection{K-12 Education (22.09\%)}
In K-12 education, which refers to primary and secondary education, ChatGPT has been used to enhance teaching and learning experiences for educators, parents, and students. The following are some of the key applications of ChatGPT in this context:

$\bullet$ \textit{Learning Language Skills.} ChatGPT has been instrumental in helping students improve their reading, writing, grammar, sentence-building, and conversational skills. By engaging with ChatGPT, learners can practice their language skills in a low-pressure environment, enabling them to gain confidence and fluency. An English teacher shared on Twitter: ``\textit{I've been using ChatGPT to help my students with their vocabulary and grammar. They love the interactive aspect and it has made a noticeable difference in their writing skills. Highly recommended!}''

$\bullet$ \textit{Personalized Learning Experiences.} ChatGPT can be used to create tailored learning experiences that cater to individual students' needs and interests. For instance, teachers can use ChatGPT to construct collaborative story-building exercises that help students learn creative writing skills in a fun and engaging manner. A middle school teacher shared their experience on LinkedIn:

``\textit{I've been incorporating ChatGPT into my creative writing lessons to create personalized prompts for each student based on their interests and skill level. It has not only helped them develop a deeper understanding of story structure and character development but also motivated them to push their creative boundaries. ... The improvement in their writing and their increased enthusiasm for the subject are truly remarkable.}''

Furthermore, ChatGPT can be integrated into educational online games (or game environments), promoting active learning and knowledge retention. A parent shared their experience on YouTube:

``\textit{ChatGPT has been a great addition to the educational games [on mathplayground] my son plays online. It acts as a smart and interactive learning assistant, guiding him through challenging math puzzles and providing real-time feedback on his problem-solving skills. ... The way ChatGPT adapts to his learning pace and offers customized hints has made learning more enjoyable and effective for him. He's now more motivated than ever to continue [exploring] and mastering new subjects.}''

$\bullet$ \textit{Supporting Home Learning.} ChatGPT has proven to be a valuable asset for supporting home learning by providing assistance with assignments and helping parents teach complex concepts. By using ChatGPT, parents can access a wealth of information and resources, enabling them to better support their children's educational journey. A parent shared in a tweet: ``\textit{ChatGPT has made a huge difference in our home learning experience. I feel more confident helping my daughter with her math homework because it offers step-by-step guidance and clear explanations. It's like having a personal tutor at our fingertips!}''

$\bullet$ \textit{Supporting Accessibility for All Students.} ChatGPT has the potential to empower students with disabilities and special needs by providing them with accessible learning resources. For instance, ChatGPT can help students with ADHD (Attention-Deficit/Hyperactivity Disorder) by offering concise explanations and engaging content tailored to their learning style. A special education teacher mentioned on LinkedIn:

``\textit{ChatGPT has been a game-changer in my classroom, especially for my students with ADHD. By using ChatGPT, I've been able to break down complex topics into shorter, more digestible segments that hold their attention. Additionally, the interactive nature of the platform keeps them engaged by providing instant feedback and allowing them to work at their own pace. The result is a more focused learning experience, leading to better understanding and retention of the material.}''

An educator who works with students with learning disabilities shared their experience in a Reddit post: ``\textit{Using ChatGPT, I've been able to develop personalized resources that cater to my students' unique learning needs. For example, I can create visual aids and simplified explanations that help them grasp complex concepts more easily. This has made a significant difference in their comprehension and overall progress. ... By leveraging ChatGPT's capabilities, educators can create inclusive learning environments that cater to the diverse needs of all students.}''

\textbf{Early Adopter Study Subject.} Based on the results of our investigation (refer to Figure \ref{fig:gptusageacross}), early adopters predominantly employed ChatGPT within the K-12 educational domain, concentrating on three primary subjects: \textit{foreign languages} (20.22\%), \textit{programming} (12.43\%), and \textit{mathematics} (11.03\%).

%%%
\subsubsection{Practical Skills Learning (15.28\%)}
In the realm of practical skills learning such as ``\textit{Do It Yourself (DIY)}'' communities \cite{10.1145/1753846.1754122}, ChatGPT has been utilized to support people in acquiring and refining hands-on abilities in various fields. Key applications of ChatGPT in this context include:

$\bullet$ \textit{Problem-solving Assistance.} ChatGPT has been influential in assisting learners with problem-solving tasks, such as debugging code or troubleshooting a malfunctioning device. By providing step-by-step guidance and engaging in question-and-answer interactions, ChatGPT enables users to develop a deeper understanding of the issues they face. For instance, one user mentioned on Reddit: ``\textit{I was struggling with a coding issue for hours, and ChatGPT helped me find the bug in no time. The [tool] provided clear instructions and explanations, making it easy for me to fix the problem and learn from the experience.}

$\bullet$ \textit{Personal Skill Development.} ChatGPT has proven valuable in helping users develop a wide range of practical skills by providing access to guidance on various projects, techniques, and tools. Examples include website development, cooking, learning a new language, and acquiring information about health and personal care. A freelance web developer shared on Twitter: ``\textit{ChatGPT has been an amazing resource for me as I've been learning web development. It offers tips, best practices, and even code snippets to help me polish my skills and create more professional websites.}''

$\bullet$ \textit{Supporting Children's Skill Development.} Parents and teachers can use ChatGPT to find age-appropriate projects and resources that help their children develop practical skills in various areas, such as programming, storytelling, cooking, or crafting. A parent shared on a Reddit comment: ``\textit{Thanks to ChatGPT, I've found so many awesome projects for my kids during their school breaks. It gives ideas that match their interests and even explains how to tackle tricky parts step by step. My kids have built robots, tried coding, and let their creativity run wild, all while having a blast. I love how ChatGPT makes learning fun and accessible and helps my family and kids grow their skills and confidence!}''

$\bullet$ \textit{Career Exploration.} ChatGPT can help users explore different career paths that involve practical skills, providing information about job requirements, necessary qualifications, and potential opportunities. Users can ask tailored questions like, ``\textit{What does it take to become an AI specialist?}'' or ``\textit{How can I excel in a web development career?}'' A college student shared on LinkedIn: ``\textit{ChatGPT has been an enlightening guide as I navigate the maze of career paths connected to my major. By shedding light on the practical skills and qualifications unique to each field, I've gained a clearer understanding of my options, ... empowering me to make well-informed decisions about my time. ChatGPT reveals insights that often remain hidden beyond the pages of academic textbooks, making it an invaluable ally in the quest for career exploration and skill development.}''

\textbf{Early Adopter Learning Areas.} As shown in Figure \ref{fig:gptusageacross}, our analysis reveals that within the realm of practical skills learning, early adopters have predominantly harnessed the power of ChatGPT to enrich their knowledge and expertise in the areas of \textit{programming and coding} (17.61\%), \textit{health and self-care} (14.28\%), and \textit{web design and development} (14.03\%).
\subsubsection{Professional Development (9.61\%)}
In the context of professional development education, ChatGPT has been employed to enhance a variety of skills and knowledge across multiple sectors. From improving communication abilities to facilitating workshop organization, ChatGPT has demonstrated its value in numerous professional settings. 

$\bullet$ \textit{Communication Skills.} ChatGPT has so far played a noteworthy role in empowering professionals to refine their written and verbal communication skills, thereby amplifying their effectiveness in their respective careers. By providing tailored suggestions and adjustments, ChatGPT guides individuals in the construction of well-organized emails, reports, proposals, and presentations. A project manager shared their experience on LinkedIn:

``\textit{ChatGPT has changed how I handle communication, giving me confidence in my writing and presenting skills. With its smart tips and on-the-spot feedback, I can write content that's not just clear and to-the-point, but also more engaging and convincing. ... Thanks to [ChatGPT], my communication game has leveled up, and it's really made a difference in how I connect with people at work.}''

By offering natural language explanations of trends, patterns, and correlations, ChatGPT simplifies the process of interpreting and communicating data-driven findings. A data researcher working at Microsoft mentioned on LinkedIn:

``\textit{[ChatGPT] not only helps me identify key insights from large datasets but also communicates those findings in a clear and concise manner that is easily understood by my colleagues and clients. It's like having an AI-powered data storytelling partner on my team!}''

$\bullet$ \textit{Digital Literacy.} Equipping individuals with digital literacy skills is critical in today's technology-driven world. Reportedly, ChatGPT has been employed by some industry sectors to facilitate learning in various areas of digital literacy, such as software usage, data analysis and management, cybersecurity, and online communication etiquette for employees. A digital marketing specialist shared on LinkedIn: ``\textit{ChatGPT has proven to be a valuable tool for my professional development. It enables me to rapidly understand new concepts, while simultaneously enhancing my decision-making abilities for my clients' benefit. It's akin to having a personal digital literacy tutor, accessible 24/7 for tailored guidance and support.}''

$\bullet$ \textit{Problem-solving.} ChatGPT has been used to boost problem-solving skills in various professional settings. By asking questions and exploring different solutions, users can work with ChatGPT to understand complex problems and come up with creative ideas. A business analyst mentioned in a Reddit post: ``\textit{Lately, when I'm stuck with a tough problem at work, I turn to ChatGPT for help. It gets me to think outside the box and see other viewpoints that I might not have thought of. The interactive and collaborative vibe of ChatGPT has really improved my problem-solving skills and overall job performance and happiness.}''

$\bullet$ \textit{Workshop Organization.} Additionally, some users have mentioned that ChatGPT has been a helpful tool in organizing and facilitating workshops, training sessions, and seminars. By creating agendas, activity plans, and resource materials, ChatGPT simplifies the planning process, letting facilitators concentrate on providing engaging and effective learning experiences. A corporate trainer shared their experience on YouTube:

``\textit{ChatGPT has been quite useful in organizing workshops for my clients. For instance, while brainstorming activities, it suggested a hands-on group exercise to encourage creative thinking, which was well-received by participants. It also assisted in creating detailed agendas and training materials, reducing my preparation time and allowing me to concentrate on delivering high-quality content. ... The feedback from participants has been predominantly positive, and I believe a significant portion of that success can be attributed to the support I've received from [ChatGPT].}''

\textbf{Early Adopter Learning Areas.} As depicted in Figure \ref{fig:gptusageacross}, our findings demonstrate that, in the context of professional development, early users have mainly utilized ChatGPT to enhance their understanding and proficiency in the fields of \textit{Information Technology (IT) and Digital Literacy} (17.08\%), \textit{Design Thinking}\footnote{As a problem-solving framework \cite{culen2014innovation}} (7.29\%), and \textit{Business Writing Skills} (6.96\%). 
\subsubsection{Entrepreneurship Education and Apprenticeship Programs (7.18\%)}
In the domain of entrepreneurial skills and expertise, ChatGPT has emerged as an invaluable information resource for entrepreneurs and startups seeking to innovate, develop, and expand their businesses. Spanning across brainstorming, benchmarking, and proposal writing, ChatGPT has reportedly played a beneficial role in numerous facets of the entrepreneurial journey.

$\bullet$ \textit{Brainstorming.} ChatGPT has demonstrated its usefulness for entrepreneurs by generating innovative ideas and presenting various viewpoints. It aids the ideation process by suggesting enhancements, pinpointing potential issues, and fostering creativity. A startup founder shared their experience on YouTube: ``\textit{With ChatGPT involved, it's like having an additional team member who brings a wide range of diverse ideas, enriching [our] group's creative thinking. ... ChatGPT has been a helpful addition to our brainstorming sessions, offering valuable insights and [alternative] approaches that led to positive, profitable outcomes. For example, [it] once suggested an interesting marketing strategy targeting a previously neglected customer segment [in the user feedback pool], resulting in success. ... Although we've seen encouraging results, we understand the need for ongoing evaluation and tracking its performance over time.}''

$\bullet$ \textit{Benchmarking.} ChatGPT can help entrepreneurs and students gain insights into industry trends, establish benchmarks, and determine essential performance indicators for comparative analysis. While the information supplied may occasionally be somewhat dated owing to ChatGPT's training data constraints, the AI assistant remains capable of delivering valuable insights into prevailing industry standards and effective benchmarking practices. A co-founder of a tech startup shared on LinkedIn: 

``\textit{Using ChatGPT, we've been able to identify crucial benchmarks and performance metrics specific to the e-commerce industry. ... [Among its contributions,] ChatGPT assisted us discover innovative strategies for reducing cart abandonment rates, which we hadn't considered before. While some information may occasionally be less relevant, ChatGPT consistently provides valuable insights that enhance our benchmarking comparisons, enabling us to gain a deeper understanding of our market position and make data-driven strategic decisions with increased confidence.}''

$\bullet$ \textit{Proposal Writing.} ChatGPT has been utilized by entrepreneurs and their teams to develop convincing grant proposals and pitch decks. By offering support in conducting literature reviews, proposing enhancements in structure, and recommending pertinent (though possibly not the most recent) references, ChatGPT assists users in crafting compelling content that appeals to potential investors and stakeholders. A grant writer for a social enterprise shared their experience on YouTube: ``\textit{It's awesome to find tools like ChatGPT that make grant proposal writing easier. It's great at finding relevant literature and organizing content, which saves a ton of effort. But its real strength is creating a persuasive language that makes proposals shine. This is key for grant writing, as you gotta show why your project deserves funding. With ChatGPT, you can be sure your proposals have a better shot at success, ... even when writing in a language other than English.}''

The wide-ranging applications of ChatGPT in entrepreneurship education and apprenticeship programs emphasize its adaptability and capacity to empower future entrepreneurs and innovators. This, in turn, cultivates a more vibrant and prosperous entrepreneurial ecosystem.

\textbf{Early Adopter Learning Areas.} As illustrated in Figure \ref{fig:gptusageacross}, our research shows that within entrepreneurship education and apprenticeship programs, early adopters predominantly use ChatGPT to improve their knowledge and skills in \textit{Design Thinking} (12.06\%), \textit{Critical Thinking} (11.05\%), and \textit{Business Modeling} (9.65\%). 
\subsubsection{Lifelong Learning (4.64\%)}
In the context of lifelong learning, ChatGPT has emerged as a valuable tool that supports self-directed learning by offering a diverse range of resources and features customized to individual needs. These features include facilitating microlearning experiences, recommending customized learning paths, and providing ongoing mentoring support.

$\bullet$ \textit{Microlearning.} ChatGPT excels at delivering engaging, bite-sized learning experiences through interactive question-and-answer sessions, concise lessons, and clear explanations of a wide array of phenomena. This targeted approach enables learners to easily grasp and retain information, making it particularly beneficial for those pressed for time or looking to deepen their understanding of specific topics. 

A YouTuber, who identifies himself as a lifelong learner, shared his enthusiasm for ChatGPT on his channel: ``\textit{As someone constantly juggling tasks, I've found that ChatGPT is exceptional at quickly explaining various subjects, from Photoshop techniques and video editing tips to makeup tutorials. The ability to ask questions and receive straightforward, focused answers has made the learning experience with ChatGPT far more enjoyable than sifting through search engine results or leafing through magazines.}'' He concludes, ``\textit{Although ChatGPT may not always offer the most current information or be entirely accurate, it remains an invaluable resource for broadening one's general knowledge as well as delving deeper into specific areas of interest or questions.}''

$\bullet$ \textit{Learning Path Suggestion.} ChatGPT can recommend customized learning paths based on an individual's self-declared goals and interests, such as establishing mastery in a programming language or natural language. By providing guidance on the ``\textit{necessary learning requirements, resources, and steps,}'' ChatGPT enhances the learning experience and helps users achieve their desired outcomes. A software developer shared their experience on a tech-oriented Subreddit:

``\textit{I was eager to learn [a new programming language] but had no idea where to begin. ChatGPT showed me the ropes, pointing out key concepts, resources, and steps to [mastering] the language in a matter of weeks. By following its [tailored] recommendations, I was able to create a personalized learning plan that met [my needs] and allowed me to [advance] as a professional at my own pace.}''

$\bullet$ \textit{Continuous Mentoring.} ChatGPT serves as a valuable resource for lifelong learners by offering comprehensive support, including access to information, resources, and constructive feedback. Whether it involves refining a written piece, nurturing creative ideas, managing projects, or drafting research findings, ChatGPT effectively functions as a consistent mentor. Notably, older adults with limited literacy skills have found ChatGPT particularly inspiring, as it enables them to continue learning without feeling shy about asking basic questions. Moreover, several laborers have credited ChatGPT for motivating them to pursue vocational education, given the platform's availability and accessibility as a constant educational companion. This highlights the substantial impact ChatGPT has on diverse learner profiles, providing tailored and timely support to individuals from various backgrounds.  However, it is widely noted among users that ChatGPT's inability to remember previous contexts poses a challenge that must be addressed in order to provide more effective and personalized mentoring for lifelong learners.

A writer seeking to develop their skills with ChatGPT's assistance shared their experience on YouTube: ``\textit{[ChatGPT] provides constructive criticism and actionable advice for enhancing my work, contributing to my growth as a writer. However, its inability to remember previous interactions can be tedious, as I often need to re-establish context. Despite this limitation, I believe ChatGPT remains a valuable asset that has had a positive impact on my writing journey.}''

In summary, ChatGPT's applications in lifelong learning demonstrate its potential to empower individuals throughout their continuous growth and development.

\textbf{Early Adopter Learning Areas.} As depicted in Figure \ref{fig:gptusageacross}, our study reveals that, in the realm of lifelong (continuous) learning, early adopters chiefly employ ChatGPT to enhance their proficiency and expertise in \textit{Languages} (21.91\%), \textit{Technology Domain} (16.22\%), and \textit{Finance} (9.06\%).

\subsubsection{Popular Science Education (3.25\%)}
Popular science education plays a crucial role in making complex scientific concepts accessible and engaging for general and often non-expert audiences \cite{10.1145/3491140.3528279}. It incorporates various media forms, such as YouTube videos and Medium articles, to effectively communicate scientific ideas, including topics like quantum physics, quantum computing, or simply the water cycle. By simplifying abstract concepts and promoting scientific literacy, popular science education fosters a greater appreciation for the scientific method and its contributions to society \cite{10.1145/3491140.3528279}. According to our findings, ChatGPT has become a valuable tool in this domain, assisting content creators and educators in achieving their educational objectives.

$\bullet$ \textit{Simplifying Complex Concepts.} ChatGPT excels at breaking down intricate scientific ideas into digestible and more comprehensible explanations. By providing analogies, examples, and concise definitions, it enhances the overall understanding of scientific concepts for a broader audience. A science communicator shared their experience on LinkedIn: ``\textit{Using ChatGPT, I recently explained gravitational waves to a group of public audience members by comparing them to ripples in a pond. This comparison struck a chord with my audience, piqued their curiosity, and made the topic appear more accessible and engaging. ChatGPT's potential for making complex and abstract ideas easier to understand while maintaining a high level of accuracy has been an incredible boon for communicating complicated scientific concepts in our online seminars and classes.}''

$\bullet$ \textit{Storytelling for Education.} In addition to simplifying complex concepts, effective popular science education often relies on storytelling to captivate and inform audiences. ChatGPT's natural language processing capabilities make it well-suited to crafting engaging narratives that weave together scientific concepts and real-world applications. A YouTube science educator shared their thoughts on the platform: "\textit{Incorporating ChatGPT into my video creation process has not only helped me simplify difficult topics but also craft compelling stories that hold my audience's attention. The combination of clear explanations and engaging narratives has significantly improved the quality of my content and contributed to an increase in viewership.}"

$\bullet$ \textit{Scientific Visualization.} The visual representation of scientific concepts poses a significant challenge to scientific content creators. However, it is crucial for enhancing comprehension and engagement in popular science education. ChatGPT can aid in generating innovative ideas for graphs, web-based interactive elements, slide designs, and other visualization techniques. Recently, ChatGPT has been utilized in conjunction with generative AI tools such as \textit{Midjourney}\footnote{https://www.midjourney.com/home/} and \textit{Stable Diffusion}\footnote{https://stablediffusionweb.com/}, enabling the creation of smart prompts, and resulting in more customized image generation. These features effectively help transform complex information into more tangible and accessible visual formats. A science writer on LinkedIn shared their experience, emphasizing the value of ChatGPT in the realm of popular science education: ``\textit{Working with ChatGPT has allowed me to come up with engaging visualizations for my [research-based] articles. ... [Just recently,] in one of our collaborative attempts, it came up with a clever analogy for a complex research concept in one of my [pieces,] and I turned it into an eye-catching infographic. ... When I presented it at a public science [exhibition,] people found it both easy to understand and visually appealing.}''

\textbf{Early Adopter Learning Areas.} As illustrated in Figure \ref{fig:gptusageacross}, our research demonstrates that early adopters in popular science education predominantly utilize ChatGPT to bolster their knowledge and skills in the fields of \textit{Computer Science and Technology} (20.62\%), \textit{Astronomy} (19.50\%), and \textit{Biology} (14.18\%).

\subsubsection{General Education (Uncategorized) (13.77\%)}
While many ChatGPT applications can be readily classified into specific educational and learning domains, certain instances of its use remain uncategorized due to the lack of (sufficient) context in social media content. Since ChatGPT applications such as creative and critical content writing and editing, text summarization, and providing helpful learning suggestions have been previously discussed and covered in earlier sections, this subsection will focus on three general topics that emerge more prominently within the context of uncategorized discourses on ChatGPT's educational applications: customized curriculum design, student assessment, and fostering collaboration.

$\bullet$ \textit{Customized Curriculum Design.} ChatGPT has the potential to help educators develop customized and adaptive curricula. By understanding the specific learning objectives and desired outcomes in an educational program (or a course of study), ChatGPT can help connect the dots and recommend customized instructional resources and approaches tailored to meet different learning goals. However, a human educator would still need to guide the overall curriculum design process, ensure the curriculum meets key requirements and standards, and incorporate human judgment in determining the optimal learning paths and experiences for students. An instructional designer conveyed their experience on YouTube, stating: ``\textit{AI won't replace instructional designers, but it'll definitely change our role. With systems like ChatGPT, we can now create, or rather draft, curricula that are personalized, adaptive, and responsive to various learning objectives super quickly. Even so, human judgment stays crucial. ... [This means] instructional designers will have an even bigger responsibility as the planners of human-AI collaboration in education.}''

$\bullet$ \textit{Student Assessment.} ChatGPT's capacity to generate tailored (practice) exams and homework assignments, covering a broad range of information and skills, provides a useful resource for educators, parents, and students themselves to monitor learning progress effectively. An educator remarked on YouTube, ``\textit{Previously, the idea of designing personalized tests appeared unattainable; however, that has changed. Utilizing ChatGPT, I've created [assessments] that address multiple [cognitive] levels and incorporate relevant real-world examples, which helps my students connect with the material, [fostering] a more authentic learning experience.}''

$\bullet$ \textit{Facilitating Collaboration.} ChatGPT can play a crucial role in fostering effective communication and collaboration, especially in diverse and multilingual settings. By providing translation services and facilitating the exchange of ideas, it helps create an inclusive learning environment where students feel empowered to work together and share their insights. A user shared on Reddit, ``\textit{This is what worldwide teamwork for learning could look like in the future. With ChatGPT's ability to understand different languages, obstacles to connecting with each other disappear and opportunities to work together arise. Students who struggle to say what's really on their mind can have their ideas translated and spread, giving them a space to be heard and making co-creating new knowledge a reality.}''

Although ChatGPT shows promise across a variety of educational fields, it is important to note that in the general education category (unspecified), the majority of learning areas, specifically more than 85\%, lack precise definitions or a concentrated focus, resulting in broader discussions on education. Moving forward, we explore RQ2, examining both the positive and negative perspectives of using ChatGPT in education to provide a more complete picture of its potential benefits and challenges.

\subsection{RQ2: Promises and Perils of ChatGPT in Education}

Our analysis of social media discussions reveals that productivity (73.19\%), efficiency (63.02\%), and ethics (48.51\%) are the predominant topics related to the integration of ChatGPT into education. Individuals have expressed diverse opinions on these aspects across various social media platforms. Our findings suggest that ChatGPT can potentially yield both positive and negative outcomes in terms of productivity, efficiency, and ethics.

Early adopters of ChatGPT within the educational sector have reported mixed perceptions. Specifically, 30.62\% expressed exclusively positive views, 17.15\% exclusively negative views, 37.34\% mixed views (encompassing both positive and negative aspects), and 14.89\% maintained a neutral stance. Table \ref{tab:perceptfigtable} presents a comprehensive illustration of the distribution of individuals' perspectives concerning the implications of ChatGPT on productivity, efficiency, and adherence to ethical principles within various educational settings. In the subsequent sections, we explain these diverse perspectives on the integration of ChatGPT into education in greater detail.

\begin{table}[t!]
  \centering
  \caption{Frequency distribution of people's perceptions regarding ChatGPT's impact on \textit{productivity}, \textit{efficiency}, and \textit{ethical values} across various educational settings.}
 \includegraphics[width=\textwidth,keepaspectratio]{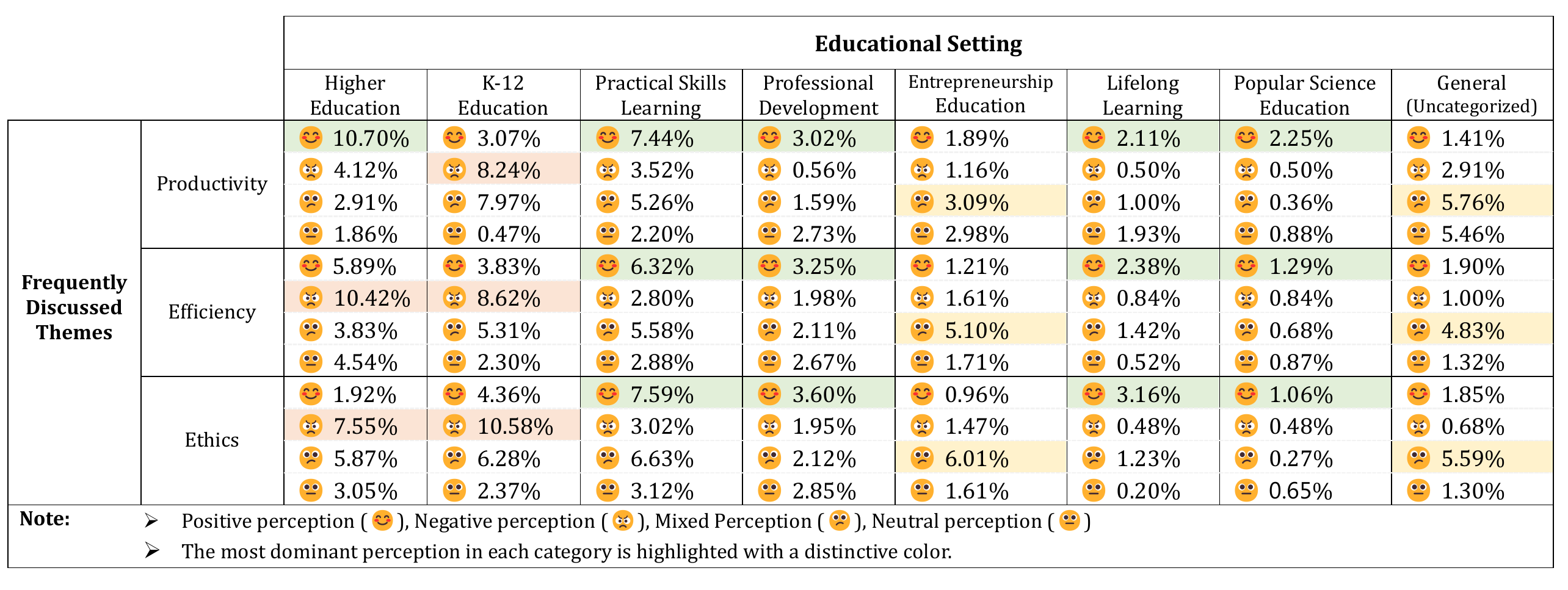}
\label{tab:perceptfigtable}
\end{table}%

\subsubsection{The Spectrum of Opinions on Productivity}
In the context of this study, productivity refers to the capacity of ChatGPT to enhance the quantity of education-related and learning outcomes within a specific timeframe. The spectrum of opinions on productivity can be divided into four primary groups:

$\bullet$ \textit{Positive Impacts on Productivity.} 
Our findings reveal that, with the exception of K-12 education, entrepreneurship, and general education (uncategorized), individuals generally possess a positive perception of ChatGPT's utility for improving productivity in the educational domain. According to our findings, these favorable views on productivity largely stem from ChatGPT's capability to streamline various educational processes, such as content creation, editing, and text summarization.

An educator's post on LinkedIn exemplifies this positive sentiment at its best: ``\textit{Ever since using ChatGPT, I've created and updated course materials much faster. It's made me more productive. For teachers who spend much personal time on admin work, this could free them to focus on students.}''

$\bullet$ \textit{Negative Impacts on Productivity.} Negative opinions are primarily found in the context of K-12 education. Teachers and parents express concern that ChatGPT may exacerbate procrastination, leading students to become increasingly disengaged from their tasks and less productive. A teacher elucidated this point on YouTube, stating, ``\textit{Lazy people may become even lazier and possibly not do anything at all. ... The thought that the task can be completed at any time can lead to doing nothing at all.}''

These concerns can be better understood in light of the well-known theory of \textit{Self-Determination} \cite{10.1145/3313831.3376723}, which posits that individuals are more motivated and engaged when they experience autonomy, competence, and relatedness in their activities. In educational settings, students who rely heavily on ChatGPT may feel a decreased sense of autonomy and competence, as the AI tool assumes most of the responsibility for their learning and academic success. This diminished sense of self-determination could lead to lower motivation and engagement, making students more passive and less productive in their education.

$\bullet$ \textit{Mixed Views on Productivity.} 
In entrepreneurship education and general education (uncategorized), opinions tend to be mixed. Users acknowledge both the benefits and drawbacks of using ChatGPT in education, suggesting that its impact on productivity depends on certain factors, most notably the prompt design and the ability to verify or organize generated content. Poorly designed prompts and inadequate domain knowledge may result in time-consuming and disorganized outcomes, which could ultimately hinder productivity.

One user on YouTube expressed this sentiment the best by sharing, ``\textit{ChatGPT can be a great help when it comes to brainstorming ideas or drafting content, potentially speeding up the process. However, you still need to invest some time refining the output and making sure it's accurate. ... While it may not be a magic solution for productivity, it can be a valuable tool if used properly.}''

$\bullet$ \textit{Neutral Views on Productivity.} 
Neutral perspectives do not explicitly express a positive or negative opinion on ChatGPT's impact on productivity in education. Instead, they report potential outcomes and call for further research, experimentation, and observation to determine ChatGPT's true influence on educational productivity.

An example of a neutral view shared by a researcher on Twitter is: ``\textit{As AI writing tools and models like ChatGPT grow more sophisticated, we need more data and research to understand their full impact on productivity in education.}''

\subsubsection{The Spectrum of Opinions on Efficiency}
In the context of this study, efficiency refers to ChatGPT's ability to enhance the quality of educational processes and outcomes, including improvements in learning, expertise, and skill development. As with productivity, the spectrum of opinions on efficiency can be divided into four primary groups:

$\bullet$ \textit{Positive Impacts on Efficiency.} 
Our findings show that in the educational settings of practical skills learning, professional development, lifelong learning, and popular science education, most users possess positive perceptions toward the utility of ChatGPT for improving the efficiency of learning.

These positive views are often attributed to ChatGPT's capacity to provide customized feedback, generate relevant examples, and offer quick and simple answers to a wide variety of questions. These features can help enhance students' understanding of complex concepts, facilitate skill development, and support educators in delivering tailored instruction.

The statement of a self-proclaimed lifelong learner on LinkedIn illustrates this positive sentiment at its best: ``\textit{I'm already noticing my conversation skills improving. I've been using ChatGPT to learn a new language by discussing interesting everyday topics in my target language. ... I ask it to act as an expert on a topic I enjoy, and then it uses all the specific terms, vocabulary, and slang while responding in the language I'm learning. ... It's really engaging and keeps me [motivated] because I get to learn all the important words I need to be fluent in the subject!}''

Another early user on Reddit provides a similar, more detailed experience sharing: ``\textit{I've been working on a coding project and I ask [ChatGPT] all the dumb questions I'm too embarrassed to put online. It's great because it teaches at my own pace and has infinite patience. I can run solutions by it, ask if they will work, and if not, why not, and if there are any issues I haven't foreseen. I use it to brainstorm, I just start typing out ideas and it gives me input. If I don't know how to tackle a certain part of the project, I write out the problem and ask it for suggestions and pros and cons. If I don't understand a concept, I can tell it what I think the concept is and ask if I've got it right. If it explains something to me that I don't fully understand, I can ask it for examples, or I can rephrase it to the AI and it will see where I went wrong / what I missed. It's a lot better than people for answering those kinds of questions because people aren't always paying attention to what you're asking, or sometimes you have trouble communicating your thoughts in a way the other person can fully understand. ChatGPT has no problems with that.}''

$\bullet$ \textit{Negative Impacts on Efficiency.} Negative opinions regarding ChatGPT are mainly found in the contexts of higher education and K-12 education. Notably, while most early adopters in higher education believe that ChatGPT can enhance productivity, they argue that it might not be as effective in terms of learning quality, particularly for advanced, intellectual, and creative subjects.

In the K-12 context, concerns about efficiency arise due to ChatGPT's potential to impede the development of critical thinking and problem-solving skills among students. Some educators contend that students might depend too heavily on AI-generated responses, resulting in a lack of deep understanding and an inability to apply knowledge in novel situations. Excessive use of ChatGPT could also diminish interaction between students and teachers, which they deem an essential element of the learning process.

A further concern in both higher education and K-12 settings is the possibility of students not verifying the validity of the information provided by ChatGPT. As AI-generated content may not always be accurate, students could inadvertently accept incorrect information as true, negatively impacting their learning quality. 

Wrapping up, a student's comment on Reddit aptly illustrates this negative viewpoint: ``\textit{I bought GPT-4 to help me explain and work through problems in my Calculus 1 class, but it hasn't been able to solve even basic math equations that involve things like related rates, linearization, etc. It's pretty disappointing to be honest. I heard GPT-4 was connected to Wolfram Alpha now and I thought it would be reliable. It still confidently provides very, very wrong answers for almost every Calculus 1 problem I ask it to solve.}''

$\bullet$ \textit{Mixed Views on Efficiency.}
Consistent with the mixed opinions on productivity, early adopters also exhibit mixed perspectives regarding the utility of ChatGPT for efficiency in the context of entrepreneurship education and other general (uncategorized) educational domains. Users believe that ChatGPT can be both a valuable resource for enhancing learning experiences and a potential distraction, depending on how it is utilized. An apt illustration of such opinions can be found in a post by an early adopter on LinkedIn: ``\textit{ChatGPT is super helpful for quick answers and tackling tough topics, but it's not foolproof. That's why it's crucial to rely on our own thinking too. For effective learning, we need to merge the AI's input with our critical thinking and problem-solving skills. This approach helps us truly grasp the subject matter and ensures we're not solely dependent on ChatGPT.}''

Aligning with this perspective, a compelling instance from the applied educational sphere of entrepreneurship education is shared by an early adopter on YouTube: ``\textit{As a beginner, ChatGPT has really helped me learn a lot about the specific language of entrepreneurship and how to give clearer feedback to potential clients. ... However, it's important to note that it doesn't quite provide a practical understanding of [the concepts it teaches.] So, I think asking experienced professionals for advice and attending workshops to fully dive into the business world atmosphere are still necessary and cannot be fully substituted by ChatGPT.}''

$\bullet$ \textit{Neutral Views on Efficiency.}
Similar to neutral perspectives on productivity, neutral views on efficiency do not explicitly express a positive or negative opinion on ChatGPT's impact on efficiency in education. Instead, they report potential outcomes and call for further research, experimentation, and observation to determine ChatGPT's true influence on educational efficiency.

An example of a neutral view shared by a researcher on Twitter is: ``\textit{The outcome of ChatGPT on student learning hasn't been conclusively determined. Further research and input are necessary to understand its actual role and to establish an appropriate balance between human-AI interactions in learning experiences.}''

\subsubsection{The Spectrum of Opinions on Ethics}
In the context of this study, ethics refers to the moral implications and concerns associated with the use of ChatGPT in educational processes. Similar to productivity and efficiency, the spectrum of opinions on ethics can be categorized into four primary groups:

$\bullet$ \textit{Positive Impacts on Ethics.} 
In line with our previous findings on productivity and efficiency, most early ChatGPT adopters in the educational settings of practical skills learning, professional development, lifelong learning, and popular science education believe that the tool will help refine ethics in the education sector over time.

These proponents base their arguments on several key factors. Our findings indicate that some early adopters appreciate ChatGPT's potential to democratize access to education and foster equitable learning opportunities for students globally. They assert that the AI's capacity to deliver personalized learning experiences and provide instantaneous feedback can help bridge the divide between students from diverse socioeconomic backgrounds and those with varying learning needs. Among these supporters, some specifically emphasize that ChatGPT could be particularly beneficial for learners with disabilities or those requiring special education.

Furthermore, these supporters underscore the role of ChatGPT in cultivating motivation for self-education among individuals of all ages and at any time, thus empowering learners to become more independent and take charge of their own educational journey in a more confident and engaging manner.

The optimistic view is probably best reflected in a tweet from an early adopter employee, stating: ``\textit{ChatGPT makes education easy to access, personalized, and open to everyone. It helps people from different backgrounds and abilities to learn and work together, breaking down the usual barriers.}''

$\bullet$ \textit{Negative Impacts on Ethics.} 
Our investigation of ChatGPT's impact on ethics primarily highlights concerns in the context of higher education and K-12 education. The top three areas of ethical concern encompass academic integrity, the spread of misinformation and fraud, and privacy and data security issues.

One of the most significant ethical issues pertains to ChatGPT's potential to promote academic dishonesty. Specifically, ChatGPT's ability to produce contextually coherent text enables students to easily plagiarize, paraphrase, or distort content in their writing assignments and exams. This compromises academic integrity and undermines the educational value of specific assessment methods. Consequently, instructors face immense difficulty in accurately assessing students' genuine understanding and skills.

Another ethical concern involves ChatGPT's inadvertent propagation of misinformation and fraud. Although AI models like ChatGPT strive to generate relevant content, they may inadvertently produce deceptive or erroneous information due to training data limitations or inherent biases. This raises apprehension regarding the dependability of ChatGPT-generated content and its potential to mislead students, particularly in the absence of rigorous fact-verification mechanisms.

Finally, incorporating AI tools like ChatGPT into pedagogical practice necessitates a strong focus on privacy and data security. AI models generally require vast amounts of data to operate effectively. Some early adopters express concerns that without well-defined protection policies and limited information about ChatGPT's inner workings, there is a risk of exposing personal data and records of tool usage to potential breaches, unintended applications, or even misuse.

In conclusion, a grad school instructor's personal experience shared on Reddit best exemplifies some of the concerns and challenges surrounding the use of AI-generated content in academia: ``\textit{Students had a project report milestone to submit and just reading through the submissions I could see ChatGPT written all over it. My students who are international students that struggle to communicate with me in emails all started writing like native speakers in their reports. But that's not enough proof that they cheated, right? Then the fun begins. All 15 students are citing papers that don't exist. Five to six citations and references and not a single one is a real paper which gave me enough proof to fail them on the project and accuse them of using ChatGPT. These students account for 1/3 of my class who are now more or less failing the class half way through the semester. A lot of the remaining 2/3 of the class are already struggling.}''

$\bullet$ \textit{Mixed Views on Ethics.}
In line with the heterogeneous opinions about productivity and efficiency, the majority of early adopters express mixed views about the impact of ChatGPT on promoting or destroying ethical values in the educational context of entrepreneurship and in other (uncategorized) educational settings. A comment from a business educator on YouTube is a good example that illustrates the mixed views on using ChatGPT: ``\textit{ChatGPT is an exceptional tool for communication, but it shouldn't replace hard work or be used for dishonest acts like intellectual theft. ... This groundbreaking technology can help professionals who struggle to express their ideas clearly. By using ChatGPT to generate alternatives or expand on concepts, they can better share their knowledge with others. ... Sadly, some may misuse ChatGPT for unethical reasons like plagiarism or taking credit for others' work. It's important to remember that the true value of ChatGPT is to support and improve your own efforts, not to replace them.}''

$\bullet$ \textit{Neutral Views on Ethics} 
Regarding ethics, neutral views do not explicitly endorse or criticize ChatGPT's influence on ethical considerations in education, similar to the perspectives on productivity and efficiency. Based on our analysis, neutral comments predominantly take the form of questions, as people want to know what is considered plagiarism with ChatGPT and what is not, what is ethical and what is not, and what the boundaries are. In a tweet, for example, a high school student asks, ``\textit{Is using AI models like ChatGPT to assist in developing and enhancing essays and research papers considered an acceptable approach?}''
\subsection{Revisiting Ethics in the Context of AI-driven Education}
\revis{
As noted earlier, AI's potential to democratize education is promising. However, according to some critics, it raises the concern of a potential homogenization of thought, as AI models are predominantly trained on widely available data. This could lead to a subtle standardization of knowledge and thinking patterns under the guise of 'personalization.'

The possibility of academic dishonesty and misinformation dissemination challenges our traditional notions of 'knowledge' and 'learning.' When AI can generate contextually relevant and coherent text, we need to reconsider what it means to 'know' something or 'learn' something. This suggests the need to shift the focus of education from mere knowledge acquisition to fostering critical thinking, creativity, and emotional intelligence—areas where AI still falls short.

Data privacy and security concerns in AI-integrated education go beyond safeguarding data. They bring to the fore the power dynamics inherent in data ownership. Questions arise about who should own the data generated from students' learning and who should be responsible for potential misuse. This necessitates a thorough exploration of data rights and ownership in an AI context.

The perspectives offering mixed and neutral views highlight the complex relationship between AI and human agency. The ethical conundrum lies in the blurred boundaries between AI as a tool to extend human capabilities and AI as an autonomous entity capable of impacting human lives.

Moreover, the biases inherent in AI could reflect our societal prejudices, effectively holding a mirror up to societal norms \cite{Foroughi2023}. This revelation can serve as a catalyst for critical discussions about hidden biases in our daily lives and how we can challenge and address these biases. This realization underscores the importance of maintaining a ``human-in-the-loop'' approach in AI systems \cite{Calisto2021}. By actively involving humans in the decision-making processes of AI systems, we can ensure that the technology is supervised, guided, and corrected by human intelligence and ethical judgment.

To sum up this section, the ethical challenges posed by the use of AI in education provide an opportunity for societal reflection and transformation. Our response to these challenges will shape not only the future of education but also our society's trajectory in an AI-driven world. The ethical considerations of AI use in education merit ongoing, critical, and nuanced dialogue among all stakeholders.}

\section{Recommendations}
\revis{This study conducted an in-depth analysis of social media content from four platforms to investigate early adopters' use and perceptions of ChatGPT in education. Our findings suggest that users utilize ChatGPT for various reasons and perceive it as a valuable educational tool that can assist students and educators with various time-intensive tasks, such as research, problem-solving, customized feedback and language learning. However, early adopters also expressed concerns about the potential misuse, including cheating, overreliance on the AI tool for answers, indolence and superficial learning.

An essential inquiry arising in this context is the optimal utilization of ChatGPT to enhance and facilitate educational objectives. Based on the collective insights derived from research questions RQ1 and RQ2, we put forth the subsequent recommendations to promote the effective incorporation of ChatGPT in educational settings while concurrently addressing some of the relevant challenges. Figure \ref{fig:flowflow1} provides an overview of all recommendations in a flowchart.

\begin{figure*}[ht!]
    \centering
    \includegraphics[width=\textwidth,keepaspectratio]{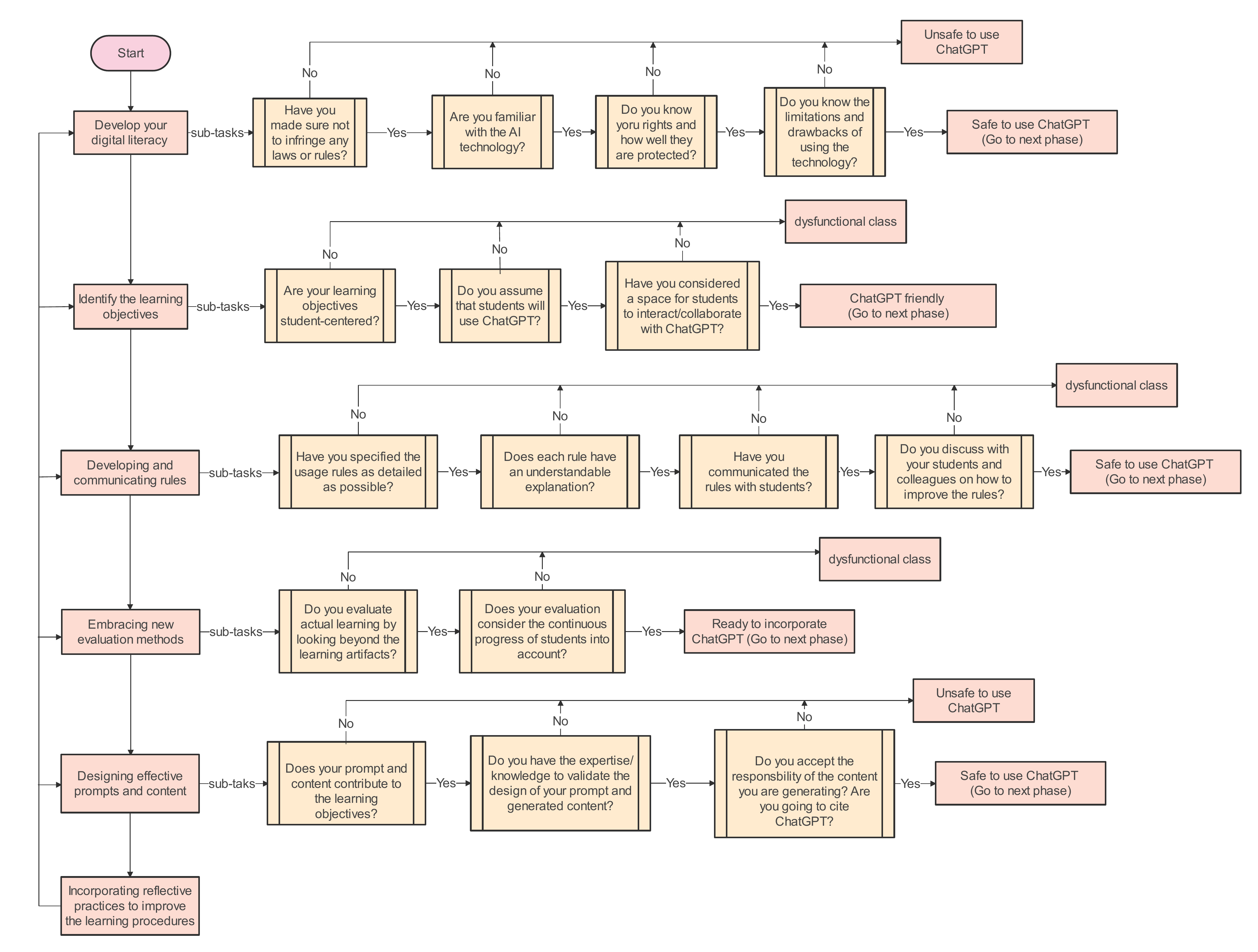}
    \caption{An overview of recommendations for integrating ChatGPT into an educational setting}
    \label{fig:flowflow1}
\end{figure*}

\subsection{Building Digital Literacy for Responsible AI Integration in Education} 
Digital literacy encompasses the competencies and skills necessary for individuals to effectively and responsibly engage with complex information ecosystems \cite{Tinmaz2022, YINOKANG2023107529}. In today's rapidly evolving educational landscape, digital literacy is essential for educators and learners to adeptly utilize technology and adopt the appropriate growth mindset \cite{Liang2021}. This is particularly significant when interacting with artificial intelligence tools, such as ChatGPT, where digital literacy is pivotal in identifying and addressing the emerging ethical, legal and societal implications \cite{Tinmaz2022, Yang2022}.

To cultivate digital literacy, grasping technology's multifaceted capabilities and fundamental mechanics is imperative. Users should have a holistic view of the underlying principles and operations of data-driven computational models and algorithms to maintain realistic expectations of tools such as ChatGPT. With this knowledge, users can discern these technologies' potential applications and limitations in educational contexts more effectively. For instance, it is noteworthy that although ChatGPT can produce coherent and contextually relevant text, it may encounter difficulties with nuanced or highly specialized subjects and technical inferences. As a result, using ChatGPT as the main reference for drafting legal documents is generally inadvisable; however, employing ChatGPT to provide an initial outline or suggestions for legal content under the supervision of legal professionals might be acceptable. Similarly, seeking medical advice from ChatGPT is generally inadvisable, but employing ChatGPT to assist patients in articulating their thoughts under the guidance of medical professionals might be acceptable. Thus, the bottom line is that users must exercise prudence when assessing the accuracy and pertinence of the generated content.

AI systems may unintentionally perpetuate and reinforce existing biases in the source material by analyzing vast but limited datasets \cite{rae2021scaling}. Hence, educators and learners must develop skills to investigate, identify and address these biases where possible. Applying such knowledge when integrating ChatGPT into learning platforms (for example, question pool websites, massive open online courses (MOOCs) and virtual learning environments) is crucial to foster a fair and inclusive learning environment. Some instructors have suggested that critical thinking, media literacy and the comprehension of the ethical implications of AI can serve as effective ways to develop these skills.

Digital literacy also encompasses the ethical dimensions of AI usage, including data privacy and security, academic integrity and the potential for biased or inappropriate content. Therefore, educators must adopt a proactive approach to ensure students know the potential risks and ethical implications of using AI tools such as ChatGPT. This involves discussing appropriate citation practices to avoid plagiarism, understanding the potential ramifications of sharing personal or sensitive information with AI systems and informing users of their rights and responsibilities regarding data ownership and usage. By doing so, learners are empowered to make informed decisions about utilizing and protecting their information.

\subsection{Reimagining Learning Goals and Objectives in the AI Era}
In the era of AI, educators must reconsider learning goals and objectives, given the likelihood that students will use AI-assisted tools such as ChatGPT for learning. New objectives should focus on developing students' abilities to collaborate with AI systems, compare AI responses, and think critically about the information provided. Activities encouraging students to debate with ChatGPT or analyze its in-depth responses could be valuable. Additionally, educators must assess the relevance and meaningfulness of their courses and lessons in an AI-infused learning environment.

As students increasingly rely on AI tools for support, educators must ensure that learning outcomes harmonize with this evolving landscape. Theoretical frameworks such as active learning \cite{prince2004does}, constructivist learning theories \cite{scheer2012transforming}, and Bloom's Taxonomy \cite{forehand2010bloom} offer valuable guidance for adapting learning goals and objectives to account for the presence of AI.

Bloom's Taxonomy is a renowned learning framework for organizing educational objectives into hierarchical levels of complexity and specificity. The taxonomy comprises the following six levels: knowledge, comprehension, application, analysis, synthesis and evaluation \cite{forehand2010bloom}. By aligning learning goals and objectives with Bloom's Taxonomy, educators can establish a structured approach to promoting cognitive development and critical thinking skills in AI-enhanced learning. Integrating AI tools such as ChatGPT into the learning process can support objectives across the various levels of Bloom's Taxonomy:

\begin{itemize}
\item \textit{Acquiring Factual Knowledge}: ChatGPT can help students establish foundational knowledge in various subjects. However, educators must instruct students on effective research methods and encourage using multiple sources for accurate understanding.

\item \textit{Understanding Complex Ideas}: ChatGPT can provide explanations, examples and analogies to support the development of comprehension skills. Herein, educators must assess and supplement AI-generated explanations for a holistic understanding (without oversimplification).

\item \textit{Applying Knowledge to Solve Problems}: ChatGPT can assist students in problem-solving, brainstorming and generating ideas in diverse subjects, fostering enhanced problem-solving skills and concept application.

\item \textit{Critically Analyzing AI-Generated Content}: Educators must encourage students to dissect AI-generated output, focusing on structure, assumptions and biases, to develop critical thinking skills. 

\item \textit{Synthesizing Diverse Information}: ChatGPT can help students combine information from different sources, promoting well-informed, balanced perspectives and a deeper understanding of knowledge interconnectedness.

\item \textit{Evaluating AI-Generated Content}: By engaging in debates with AI tools or analysing AI-generated content in depth, students can develop their evaluative reasoning skills, learning to assess the quality and credibility of information and make informed judgments.  
\end{itemize}

\subsection{Developing and Communicating Rules for AI in Education} 
Educators must develop and communicate specific rules outlining the appropriate use of ChatGPT in various settings. By providing well-reasoned explanations for each rule, teachers can foster a sense of cooperation and understanding among students \cite{McCabe2001}. For instance, a rule might state: ``\textit{Students are encouraged to use ChatGPT as a supplementary resource to support their learning and research activities, complementing their engagement with primary sources, diverse perspectives and critical thinking skills, but not as a substitute for these essential components of learning. This approach ensures that students do not excessively rely on AI and, in turn, get the space to develop robust cognitive skills necessary for independent learning.}'' 

By establishing well-defined parameters, educators can ensure that students harness the potential of AI technologies while minimizing the risks of misuse or overreliance. Well-established psychological theories, such as the Theory of Planned Behaviour and the Self-Determination Theory, can support the establishment of these rules and guidelines.

The Theory of Planned Behaviour (TPB) posits that an individual's intention to perform a specific behavior is determined by their attitude towards the behavior, subjective norms and perceived behavioral control \cite{ajzen1991theory}. Educators can influence students' attitudes and subjective norms regarding the responsible use of AI tools such as ChatGPT by defining clear rules and guidelines (especially considering the nascent stage of this technology). This approach can, in turn, shape students' intentions to use technology responsibly and effectively. Furthermore, by explaining the rationale behind each rule, educators can foster a sense of shared understanding and cooperation, strengthening the subjective norms that govern responsible AI tool usage.

Self-Determination Theory (SDT) is another psychological framework that emphasizes the importance of intrinsic motivation and autonomy in human behavior \cite{10.1145/3313831.3376723}. According to SDT, individuals are likelier to engage in a behavior when they perceive it to be autonomously chosen and intrinsically motivating. Establishing clear rules and guidelines for AI tool usage and providing well-reasoned explanations can promote a sense of autonomy and intrinsic motivation among students. Students may be more inclined to follow the guidelines and adopt responsible behaviors once they understand the purpose and benefits of responsible AI usage.

In addition to providing clear rules, maintaining open communication with students about the responsible use of AI tools is crucial. Regular discussions on the potential benefits and pitfalls of using ChatGPT in learning can promote a deeper understanding of technology's role in education. This dialogue can also facilitate the exchange of ideas, feedback and experiences, enabling educators and students to collaboratively refine best practices for effectively integrating AI tools into the learning process.

Moreover, educators should consider regularly reviewing and updating the rules and guidelines to safeguard their continued relevance and effectiveness. As AI technologies evolve and new insights emerge from research and practice, adapting the guidelines to reflect these changes will be essential for maintaining an up-to-date and responsible approach to AI integration in education.  

\subsection{Embracing New Evaluation Approaches with AI Integration}
Focusing on actual learning outcomes rather than solely on the artifacts produced by students is vital for outlining effective evaluation approaches \cite{10.1145/3491140.3528274}. In line with the extant literature, our findings show that most educators concur that continuous assessment throughout the semester facilitates a more precise evaluation of learning compared to a few high-stakes exams \cite{Nistal2013}. With the advent of ChatGPT, educators are increasingly inclined to assess students primarily within classroom settings.

Project-based assessments, which align with higher levels of thinking in Bloom's Taxonomy, are an efficient means of evaluating learning in environments employing AI tools. These assessments emphasize the application of knowledge and skills in real-world contexts, focusing on the development of higher-order cognitive processes such as analysis, evaluation and creation. By devising and executing projects that demand critical thinking, problem-solving and innovative thinking, educators can motivate students to engage with AI tools responsibly and thoughtfully, fostering the growth of crucial twenty-first-century competencies.

Moreover, incorporating peer and self-assessment techniques can enrich the evaluation process by promoting collaboration, reflection and metacognitive skills among students \cite{noonan2005peer}. These methods empower students to assess their and their peers' works, providing opportunities for constructive feedback and fostering a deeper comprehension of the assessment criteria. Nonetheless, it is crucial to acknowledge that most instructors deem focused guidance from a supervisor indispensable in such activities. Such guidance enables students to reflect more swiftly on their tasks, assignments and exams by participating in peer and self-assessment activities and utilizing AI tools such as ChatGPT. Many educators argue that this approach results in more meaningful and enduring learning outcomes.

In addition to re-envisioning assessment and evaluation approaches, educators must address the ethical implications of AI integration within learning environments. This entails confronting data privacy, algorithmic fairness and potential biases in AI-generated content. Integrating discussions of these ethical considerations into the curriculum and assessment practices enables educators to raise awareness and instill a sense of social responsibility among students when using AI tools for assignments and exams.

\subsection{Designing Effective Prompts and Incorporating Reflective Practices in AI-Assisted Education}

The successful integration of AI tools such as ChatGPT in educational settings increasingly relies on crafting well-structured prompts and incorporating reflective practices. Thoughtful prompt design enables educators to leverage ChatGPT's potential to support instructional goals and enhance student learning. Furthermore, integrating reflective practices fosters critical thinking about learning, using AI tools, and developing essential skills and competencies. 

Designing quality prompts begins with identifying and aligning clear learning objectives with appropriate AI-supported activities. This process entails evaluating ChatGPT's strengths and limitations and considering students' needs and contexts. Aligning prompts with learning goals optimizes AI tools' potential to support and enrich the learning experience.

Ensuring the accuracy, relevance and suitability of ChatGPT-generated content requires testing prompts and verifying content. Educators must thoroughly review AI-generated content and cross-check it with credible sources to maintain quality and reliability. Encouraging students to cite ChatGPT when using its content promotes accountability and ethical practices in AI tool usage. 

Educators can facilitate reflection by prompting students to consider the following questions:
\begin{itemize}
    \item How did ChatGPT contribute to my understanding of the topic?
    \item What strategies did I employ to verify the accuracy and reliability of the AI-generated content?
    \item How can I use AI tools responsibly and ethically to support my learning?
    \item What challenges did I encounter while using ChatGPT, and how can I overcome them in the future?
\end{itemize}

Lastly, educators should present ChatGPT-generated content engagingly, incorporating sound pedagogy and study approaches. Techniques such as challenging questions, problem-solving, providing explanations and storytelling foster active engagement, critical thinking and knowledge retention. For instance, educators can use ChatGPT-generated content to create case studies, simulations or role-playing activities, allowing students to apply their knowledge and skills in real-world contexts. }

\section{Limitations}
\revis{Inherent to any research, our work is subject to certain limitations that must be acknowledged when evaluating its explanatory power and generalizability. While our study provides valuable insights into the experiences of early ChatGPT adopters, it encounters typical challenges associated with qualitative research, such as unintentional researcher biases (see \cite{10.1145/3555124}) and the complexities of generalizing findings beyond the specific context of the study (see \cite{10.1145/3555124, 10.1145/3410404.3414259, 10.1145/3290605.3300698}).

A primary limitation concerns the narrow scope and scale of data collection, which focuses on a particular subset of posts, time frames, and social media platforms. Consequently, our investigation may not completely capture the myriad perspectives and experiences of ChatGPT users, necessitating a prudent interpretation of our findings when drawing broader conclusions about the larger population of early adopters.

Another limitation to acknowledge is that our study does not purport to encompass the perspectives of every demographic within the early ChatGPT adopters in education. Instead, it represents the opinions of a specific group of English-speaking users who have chosen to share their experiences on the social media platforms investigated in this research. This nuance is vital, as not all users may be inclined to discuss their experiences and viewpoints online (see \cite{10.1145/3414841, 10.1145/3555124}).

To address some of these limitations and advance our understanding of user experiences and viewpoints, future research could employ a more comprehensive methodological approach. By incorporating techniques such as Delphi studies, interviews, and surveys, researchers can gather data from a larger and more diverse sample, thus providing a more robust representation of the perspectives and experiences of early ChatGPT adopters in education.}

\section{Future Work}
\revis{As the nascent field of ChatGPT studies progresses, there is a wealth of research opportunities that warrant further exploration to either corroborate or challenge the findings of our preliminary investigations. Based on the data collected and scrutinized in our research on the use of ChatGPT in education on various social media platforms, we highlight several avenues for future scholarly inquiries.

Regarding ChatGPT's practical performance, it is also crucial to assess its effectiveness in relation to specific learning outcomes, such as knowledge retention, skill development, and other pertinent objectives. Controlled experiments can be employed to facilitate this evaluation and determine ChatGPT's overall efficacy within diverse and specialized educational contexts. Concurrently, examining interpersonal dynamics is essential to gauge the impact of ChatGPT on teacher-student and student-student relationships, interaction patterns, and communication modes. This research area should explore ChatGPT's potential to promote peer learning and foster collaborative educational environments.

A critical perspective entails examining the longitudinal effects of ChatGPT on learning outcomes and users' perceptions. In-depth, longitudinal investigations can provide a holistic understanding of ChatGPT's enduring influence on education. Concurrently, a comparative analysis should be undertaken to evaluate ChatGPT against other AI chatbots and traditional pedagogical methods. This comparative framework will enable a comprehensive examination of the unique advantages and challenges associated with ChatGPT and alternative educational approaches.

The ethical and responsible integration of ChatGPT in educational settings necessitates an examination of policy development and teacher training. Policymakers, including academic institutions, must devise and implement regulations to ensure the judicious use of ChatGPT. Furthermore, it is crucial to evaluate strategies for equipping educators with the necessary skills and knowledge to seamlessly incorporate ChatGPT into their teaching practices.

Addressing the varied needs of learners, research should investigate ChatGPT's ability to cater to culturally and linguistically diverse students, as well as its application in special education. This line of inquiry should evaluate potential synergies between ChatGPT and other assistive technologies to support students with special educational needs, learning disabilities, or other distinct learning challenges.

Finally, it is essential to explore innovative approaches to enrich the learning experience with ChatGPT. One such approach involves utilizing affective computing to assess ChatGPT's ability to identify and respond to the emotional and social dimensions of learning, thereby catering to learners' affective needs. Additionally, integrating ChatGPT with various instructional strategies, such as game-based learning, gamification techniques, blended learning, and multimodal content delivery, can serve to enhance student motivation, engagement, and learning outcomes.

Moreover, the potential of combining ChatGPT with metaverse-based classrooms should be examined. Metaverse-based classrooms (see \cite{zhou2023unleasing}), utilizing immersive virtual environments, offer opportunities for transformative pedagogical experiences that foster deeper learning and collaboration. Investigating the synergy between ChatGPT and metaverse-enhanced learning environments can provide valuable insights into the effectiveness of these innovative approaches in promoting a more engaging and enriching educational experience for learners.}

\section{Conclusion} \label{sec:conclusion}
In this study, we investigated the adoption and perception of ChatGPT in education by analyzing qualitative data collected from various social media platforms. Our analysis revealed that ChatGPT is utilized in diverse settings for multiple purposes, with the most widespread uses observed in higher education, K-12 education, and practical skills development.

Among the numerous applications of ChatGPT in higher education, content creation and editing emerged as the most prevalent, accounting for 78.11\% of reported uses. Our findings unveiled a range of promising opportunities for ChatGPT to support learning, while simultaneously uncovering critical risks associated with overreliance on the tool. These risks encompass the potential to limit critical thinking and creativity, impede a deep understanding of subject matter, and foster laziness and passivity.

Drawing upon insights gleaned from the abundance of crowdsourced knowledge on social media platforms, our study offered recommendations for a more responsible and effective integration of ChatGPT in educational contexts. Moving forward, it is clear that ongoing research is crucial for tracking the progress of language models and AI tools like ChatGPT, as well as assessing their influence on learning outcomes. In accordance with this perspective, we concluded our paper by outlining an extensive research agenda for future exploration.

\newpage
% \section*{References}
\bibliography{ref}

\end{document}